\newcommand{\teff}{T$_{\mathrm{eff}}$}

\newcommand{\logg}{log $g$}

\newcommand{\bprp}{$G_{BP}-G_{RP}$}

\newcommand{\project}[1]{\textsl{#1}}
\newcommand{\gaia}{\project{Gaia}}

\documentclass[a4paper,fuseAMS,leqn,usenatbib]{mnras}

\usepackage{graphicx}
\usepackage{amssymb}
\usepackage{amsmath}
\usepackage{times}
\usepackage{subfigure}
\usepackage[T1]{fontenc}
\usepackage{ae,aecompl}
\usepackage{textcomp}

\usepackage{hyperref}
\title[CEMP in \gaia\ DR3]{Carbon-Enhanced Metal-Poor star candidates from BP/RP Spectra in \gaia\ DR3}

 \author[Lucey, et al. 2022]{Madeline~Lucey$^{1}$\thanks{E-mail:m\_lucey@utexas.edu},
 Nariman Al Kharusi$^{1}$,
 Keith Hawkins$^{1}$,
 Yuan-Sen Ting$^{2,3}$,
  \newauthor
 Nesar Ramachandra$^{4,5}$,
 Adrian M. Price-Whelan$^{6}$,
 Timothy C. Beers$^{7,8}$,
  \newauthor
 Young Sun Lee$^{9}$,
 and
 Jinmi Yoon$^{10,8}$
\\
$^{1}$Department of Astronomy, The University of Texas at Austin, 2515 Speedway Boulevard, Austin, TX 78712, USA \\
$^{2}$Research School of Astronomy \& Astrophysics, Australian National University, Cotter Road, Weston Creek, ACT 2611, Canberra, Australia \\
$^{3}$Research School of Computer Science, Australian National University, Acton ACT 2601, Australia \\
$^{4}$Computational Science Division, Argonne National  Laboratory, 9700 South Cass Avenue, Lemont, IL 60439, USA \\
$^{5}$High Energy Physics Division, Argonne National Laboratory, 9700 South Cass Avenue, Lemont, IL 60439, USA \\
$^{6}$Center for Computational Astrophysics, Flatiron Institute, 162 5th Avenue, New York, NY 10010, USA \\
$^{7}$Department of Physics and Astronomy, University of Notre Dame, Notre Dame, IN. 46556, USA \\
$^{8}$JINA Center for the Evolution of the Elements (JINA-CEE), USA\\
$^{9}$Department of Astronomy and Space Science, Chungnam National University, Daejeon 34134, Republic of Korea \\
$^{10}$Space Telescope Science Institute, 3700 San Martin Dr., Baltimore, MD 21218, USA
}

\date{Accepted . Received ; in original form }

\pubyear{2021}
\begin{document}
\label{firstpage}
\pagerange{\pageref{firstpage}--\pageref{lastpage}}
\maketitle

\begin{abstract}
Carbon-enhanced metal-poor (CEMP) stars comprise almost a third of stars with [Fe/H] < --2, although their origins are still poorly understood. It is highly likely that one sub-class (CEMP-$s$ stars) is tied to mass-transfer events in binary stars, while another sub-class (CEMP-no stars) are enriched by the nucleosynthetic yields of the first generations of stars. Previous studies of CEMP stars have primarily concentrated on the Galactic halo, but more recently they have also been detected in the thick disk and bulge components of the Milky Way. \textit{Gaia} DR3 has provided an unprecedented sample of over 200 million low-resolution ($R\approx$ 50) spectra from the BP and RP photometers. Training on the CEMP catalog from the SDSS/SEGUE database, we use \texttt{XGBoost} to identify the largest all-sky sample of CEMP candidate stars to date.  In total, we find 58,872 CEMP star candidates, with an estimated contamination rate of 12\%. When comparing to literature high-resolution catalogs, we positively identify 60-68\% of the CEMP stars in the data, validating our results and indicating a high completeness rate. Our final catalog of CEMP candidates spans from the inner to outer Milky Way, with distances as close as $r \sim$ 0.8 kpc from the Galactic center, and as far as $r >$ 30 kpc. Future higher-resolution spectroscopic follow-up of these candidates will provide validations of their classification and enable investigations of the frequency of CEMP-$s$ and CEMP-no stars throughout the Galaxy, to further constrain the nature of their progenitors.

\end{abstract}


\section{Introduction}

\label{sec:Introduction}

Stellar chemical abundances act as a fossil record of the interstellar medium (ISM) from the time a star is formed, given that a star's atmospheric abundances are not expected to change over its lifetime, aside from mass-transfer events and evolutionary changes on the giant branch.  Therefore, it is generally true that more metal-poor stars formed at earlier times when the Universe contained fewer metals. Furthermore, the detailed chemical composition of metal-poor stars can illuminate the early chemical evolution of the Universe that resulted from the lives and deaths of the first generations of stars. 

Many studies of metal-poor stars have focused on the Galactic halo, where the metallicity distribution function is dominated by metal-poor stars \citep[e.g.,][]{Beers2005,Frebel2015}. These studies have found a significant fraction of metal-poor stars in the halo that exhibit high levels of carbon enhancement ([C/Fe] $>$ +0.7), and are referred to as carbon-enhanced metal-poor (CEMP) stars \citep[e.g.,][]{Beers1992,Beers2005,Christlieb2008}. This fraction increases with decreasing metallicity in the Galactic halo, with CEMP stars making up 10--30\% of stars with [Fe/H] < --2 and $\approx$ 80\% of stars with [Fe/H] < --4 \citep{Lucatello2006,Lee2013,Placco2014,Yoon2018}. However, as cautioned by \citet{Arentsen2022}, it can be difficult to compare these fractions across different samples of CEMP stars, given the various selection effects and differences in the abundance analysis from study to study.

Further analysis of CEMP stars have identified a number of sub-classes \citep{Beers2005}. A significant fraction of CEMP stars exhibit enhancements of slow neutron-capture ($s$-process) elements (such as Ba), and are thus called CEMP-$s$ stars. There  also exist small numbers of CEMP-$r$ stars, which show enhancements in rapid neutron-capture ($r$-process) elements (such as Eu), CEMP-$r$/$s$ stars, which exhibit enhancements in both $r$- and $s$-process elements \citep{Gull2018}, and CEMP-$i$ stars, which exhibit enhancements of intermediate neutron-capture ($i$-process) elements \citep{Frebel2018}. The CEMP-no sub-class of stars does not exhibit over-abundances of neutron-capture elements.  The CEMP-$r$, CEMP-$r$/$s$, and CEMP-$i$ sub-classes are sparsely populated in extant samples, while the CEMP-$s$ and CEMP-no stars are the most common (see, e.g., \citealt{zepeda2023}).

It is thought that CEMP-$s$ stars, which are more common at [Fe/H] $>$ --3.0, are the result of chemical enrichment by mass-transfer events from (post-)asymptotic giant branch (AGB) stars \citep{Lugaro2012,Placco2013}. This hypothesis is strongly supported by the high rate of binarity among CEMP-$s$ stars \citep{McClure1990,Preston2001,Lucatello2005,Bisterzo2010,Abate2015,Hansen2016,Jorissen2016}. In fact, binarity rates as high as 82\% have been reported for CEMP-$s$ stars \citep{Hansen2016}. 

On the other hand, CEMP-no stars have a lower rate of binarity than CEMP-$s$ stars, thus are less likely to have experienced a mass-transfer event \citep{Starkenburg2014,Hansen2016a,Arentsen2019}. Hence, CEMP-no stars likely formed from an ISM that was already carbon enhanced. Given their low metallicity and increasing frequency at lower metallicities, it is thought that these are truly ancient, and were primarily enriched by the first generations of stars. It has been suggested that massive first stars may have had high rotation rates, which would lead to large carbon production \citep{Chiappini2006,Meynet2006}. Furthermore, it is possible that the first stars exploded as faint supernovae, which also overproduce carbon \citep{Umeda2003,Nomoto2013,Tominaga2014}. \citet{Yoon2016} have associated CEMP-no stars with their Morphological Groups III and II, respectively, corresponding to these two primary carbon-production sources.

Initial studies indicate that the frequency of CEMP-$s$ and CEMP-no stars varies throughout the Galaxy. Specifically, the number of CEMP stars appears to increase with increasing distance from the Sun, although we note that these studies are mostly focused on the Galactic halo \citep{Frebel2006,Carollo2012,Lee2017,Yoon2018}. Furthermore, the relative fraction of CEMP-no stars compared to CEMP-$s$ stars also increases at larger distances \citep{Carollo2014,Lee2019}. Ultra-faint dwarf galaxies have shown similar fractions of CEMP-no stars as the Milky Way, but dwarf spheroidal galaxies have a clear deficit of CEMP-no stars \citep{Norris2010,Lai2011,Frebel2014,Skuladottir2015a,Salvadori2015}. In a comparative study between Galactic halo and dwarf galaxy CEMP stars, \citet{Yoon2019} suggests that the majority of Galactic halo CEMP-no stars have been accreted from dwarf galaxies. Furthermore, CEMP-no and CEMP-$s$ stars have been discovered in the metal-weak thick disk \citep[MWTD;][]{Beers2017}.  \citet{Dietz2021} tentatively associated the retrograde MWTD CEMP-no population with the Gaia-Enceladus system, while suggesting that the equivalent prograde population has both in-situ and ex-situ origins. 

There are fewer studies of metal-poor stars towards the center of the Galaxy compared to the Galactic halo. This is partly due to the difficulty of targeting metal-poor stars in a region of the Galaxy that is dense with metal-rich stars. Furthermore, high levels of extinction demand long exposure times and large-aperture telescopes in order to achieve sufficient signal-to-noise spectroscopic observations for metallicity measurements. Fortunately, the advent of metallicity-sensitive photometric surveys \citep[e.g., Skymapper and Pristine;][]{Bessell2011,Wolf2018, Starkenburg2017b} have led to studies of thousands of metal-poor inner Galaxy stars. Studies using SkyMapper photometry have found a much lower fraction of CEMP stars in the inner Galaxy compared to the Galactic halo \citep{Howes2014,Howes2015,Howes2016,Lucey2022}. However, metallicity estimates from Skymapper photometry have proven to be biased against CEMP stars \citep{DaCosta2019,Chiti2020}. Targeting with Pristine photometry, \citet{Arentsen2021} found a CEMP frequency that is consistent with the Galactic halo for stars with [Fe/H] <--3, but also found that it is much lower than the halo at higher metallicities. 

Measuring and understanding the frequency and relative rates of CEMP-no/CEMP-$s$ stars throughout the Galaxy will be crucial for shedding new light on the origins and formation mechanisms of these stars, including whether or not CEMP-no stars are true inheritors of the elements created by the first stars. Given that the measured properties of CEMP samples have been shown to vary across different samples \citep{Arentsen2022}, creating a uniformly analyzed sample with limited selection effects across the Milky Way will be essential for achieving this goal. The release of the \textit{Gaia} BP/RP spectra in DR3 presents a unique opportunity to identify the largest all-sky sample of CEMP stars to date \citep{Witten2022}. However, the BP/RP spectra have quite low resolution ($R = \lambda/\Delta \lambda$ $\approx$ 50), and require unconventional methods for analysis.

In this work, we present a novel method for detecting CEMP stars in the \textit{Gaia} BP/RP spectra with machine learning, specifically the \texttt{XGBoost} classification algorithm, and apply it to the spectra released in DR3. In Section \ref{sec:data} we describe the BP/RP spectra, along with other data used in our analysis. We introduce \texttt{XGBoost}, our chosen classification algorithm, in Section \ref{sec:xgb}. We evaluate the accuracy and sensitivity (i.e., completeness) of our classification in Section \ref{sec:p&c}. In Section \ref{sec:interp}, we interpret the \texttt{XGBoost} model. Finally, in Section \ref{sec:sample} we present the sample of CEMP candidate stars, along with an investigation of their metallicity and Galactic distributions. 

\section{Data} \label{sec:data}

\begin{figure*}

    \includegraphics[width=\linewidth]{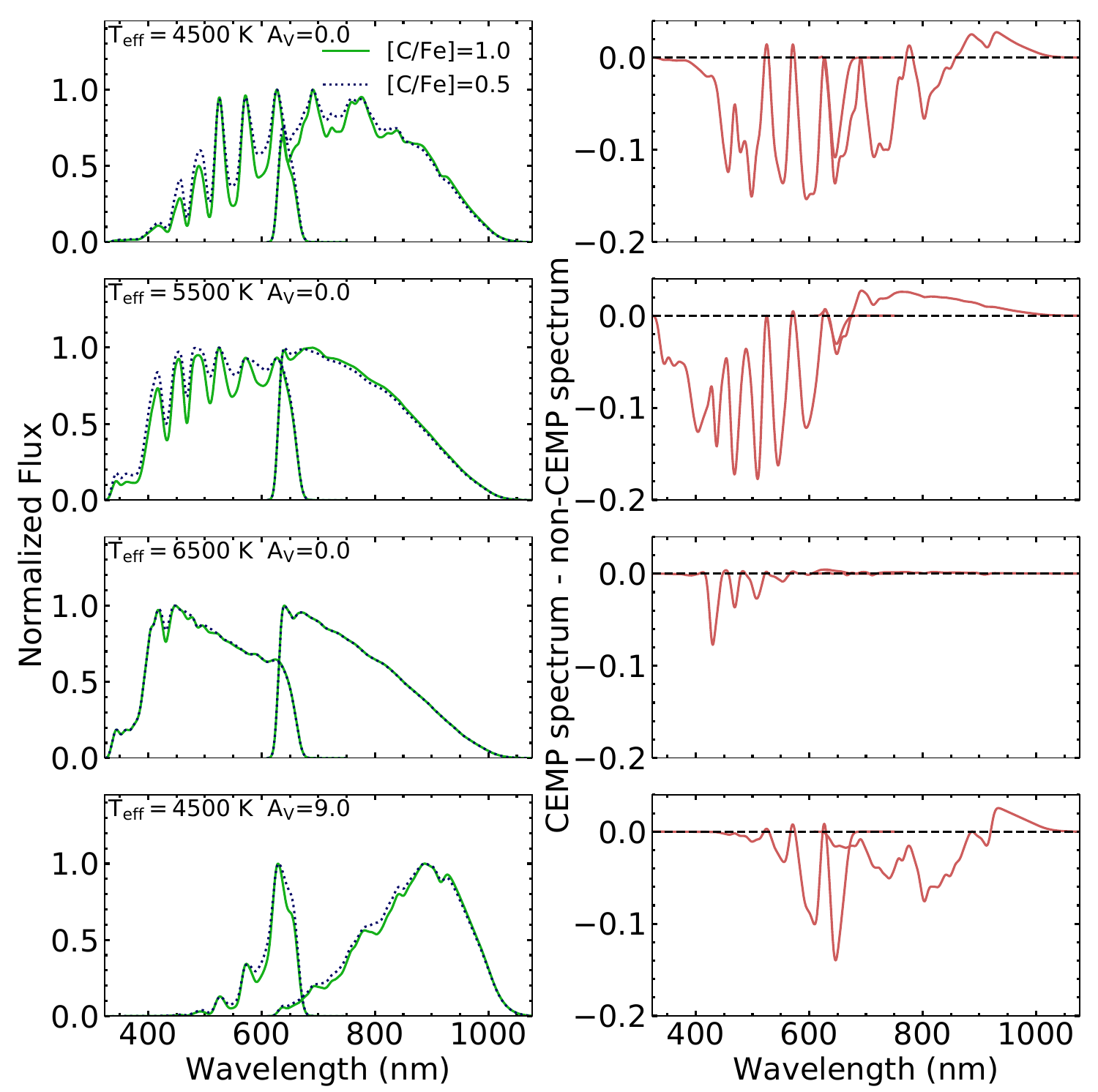}
    \caption{Examples of noiseless synthetic mock BP and RP spectra (left), and the impact of carbon enhancement on the spectra (right). Specifically, in the left panels, we compare stars of the same stellar parameters and reddening, except with different carbon abundances. The BP and RP spectra are plotted separately, with the BP spectra at lower wavelengths. The BP and RP spectra overlap at $\approx$650 nm. The blue dotted lines have [C/Fe] = +0.5, while the green solid lines have [C/Fe] = +1.0. Starting from the top, which has typical stellar parameters for a metal-poor giant (\teff = 4500\,K, \logg\ = 2.5 and [Fe/H] = --2.0), we increase the \teff\ to 5500\,K in the second row and 6500\,K in the third row. The fourth row has \teff = 4500\,K, but with increased extinction at $\rm{A_V}$=9.0 mag.  In the right panels, we have subtracted the dotted spectrum ([C/Fe] = +0.5) from the solid spectrum ([C/Fe] = +1.0) for each row. The impact of carbon on the spectra changes drastically with the stellar parameters with higher \teff s having weaker signals and extinction erasing the signal in the bluest wavelengths. We therefore require a flexible classification model in order to achieve low contamination of our detected carbon-enhanced stars. }
    \label{fig:spec}
\end{figure*}

The \gaia\ mission has revolutionised Milky Way astronomy and beyond, primarily by providing astrometric data for billions of stars \citep{Gaia2016,DR3}. Simultaneously, the \gaia\ mission has also been collecting low-resolution spectra ($R \approx$ 50), with the blue photometer (BP) and red photometer \citep[RP;][]{DeAngeli2022}. These spectra have provided effective temperature (\teff), surface gravity (\logg), and metallicity ([M/H]) estimates \citep{Liu2012,Andrae2022}, but have too low of a resolution to provide further elemental abundances \citep{Gavel2021}. Molecules, however, absorb large bands of light, and therefore may be easier to detect in the BP/RP spectra. Given the wavelength coverage of 3300--10500\,{\AA}  \citep{Carrasco2021}, we expect to be able to detect carbon-enhanced stars from the plethora of carbon molecular bands in that range (e.g., $\rm{C_2}$ Swan bands at $\approx$ 4500--6000\,{\AA}, and CN bands at $\approx$ 7000--10500\,{\AA}).  The ability to measure carbon abundances from mock BP/RP spectra has been explored by \citet{Witten2022}, but they make use of a different method than this work and also do not model the impact of dust extinction.

To explore the impact of carbon on the BP/RP spectra, we create mock synthetic spectra. We employ the MARCS carbon-enhanced model atmosphere grids \citep{Gustafsson2008} and the TURBOSPECTRUM radiative transfer code \citep{Alvarez1998,Plez2012} to construct these spectra. We also use the fifth version of the Gaia-ESO atomic linelist \citep{Heiter2020} with the addition of molecular lines for CH \citep{Masseron2014}, CN, NH, OH, MgH, $\rm{C_2}$ (T. Masseron, private communication), SiH (Kurucz  linelists\footnote{\url{http://kurucz.harvard.edu}}), and  TiO,  ZrO,  FeH,  CaH  (B. Plez,  private communication). To apply the instrumental profile of the BP/RP photometers, we use the DR3 calibrated passbands \citep{Riello2021}. The true spectral resolution is a function of wavelength for both BP and RP spectra, with the BP spectra ranging from $R \approx$ 100 at the blue edge to $R \approx$ 30 at the red edge, and the RP spectra ranging from $R \approx$ 100 at the blue edge to $R \approx$ 70 at the red edge. To simplify the calculation, we assume a resolution of 50 for the BP spectra and 70 for the RP spectra. We also model the impact of dust extinction using the \texttt{DUST\_EXTINCTION}\footnote{\url{https://github.com/karllark/dust_extinction}} package with the extinction curve from \citet{Fitzpatrick2004}, assuming $R_V$=3.1. 

Figure \ref{fig:spec} shows examples for a number of noiseless synthetic mock spectra, along with the impact of carbon on the calculated flux. The left panels shows eight spectra, in sets of two with the same stellar parameters, except one spectrum has [C/Fe] = +0.5 (blue dotted line) and the other has [C/Fe] = +1.0 (green solid line). We start with the stellar parameters of a typical metal-poor giant star (\teff\ = 4500\,K, \logg = 2.5 and [Fe/H] = --2.0) in the top row. We then increase the \teff\ to 5500\,K in the second row and \teff\ = 6500\,K in the third row. The spectra in the bottom row have \teff\ = 4500\,K but $\rm{A_V}$ = 9.0. This is a worst-case example, since we expect very few spectra released in \gaia\ DR3 to have $\rm{A_V}$ $\geq$ 9.0, given that most stars have $G < 17.6$. In the right panels, we subtract the spectrum with [C/Fe] = +1.0 from the spectrum with the same stellar parameters, but with [C/Fe] = +0.5. 

The impact of carbon enhancement on the BP/RP spectra (Fig.~\ref{fig:spec}) is very dependent on the \teff\ and reddening. Compared to the standard metal-poor giant (\teff\ =4500\,K), the hottest star (\teff\ = 6500\,K) has a significantly weaker signal.  This is likely a consequence of the dissociation of carbon molecules in the atmospheres of hotter stars (\teff > 6000\,K). This is consistent with results from \citet{Witten2022}, which found that the \teff\ must be $<$ 6000\,K in order to achieve precision on the carbon abundance of $<$ +0.5  for stars with [Fe/H] = --2.0 with \gaia\ BP/RP spectra. Furthermore, reddening greatly reduces the flux, and therefore the strength of the  carbon-enhancement signal in the bluest wavelengths. Fortunately, there are carbon molecules (e.g., CN bands at $\approx$ 7000--10500\,{\AA}) that impact the RP spectra, so we are still able to detect reddened carbon-enhanced stars. These four combinations of \teff\ and reddening values lead to significantly different results for the impact of carbon enhancement on the BP/RP spectra. As we hope to be able to detect carbon-enhanced stars across a wide range of stellar parameters and reddening, we require a complex model that can adapt to the different signals of carbon enhancement. To balance complexity with interpretability, we select \texttt{XGBoost} as our algorithm for detecting carbon-enhanced stars.

\subsection{Training and Testing Sample}
\label{sec:train}

\begin{figure*}

    \includegraphics[width=\linewidth]{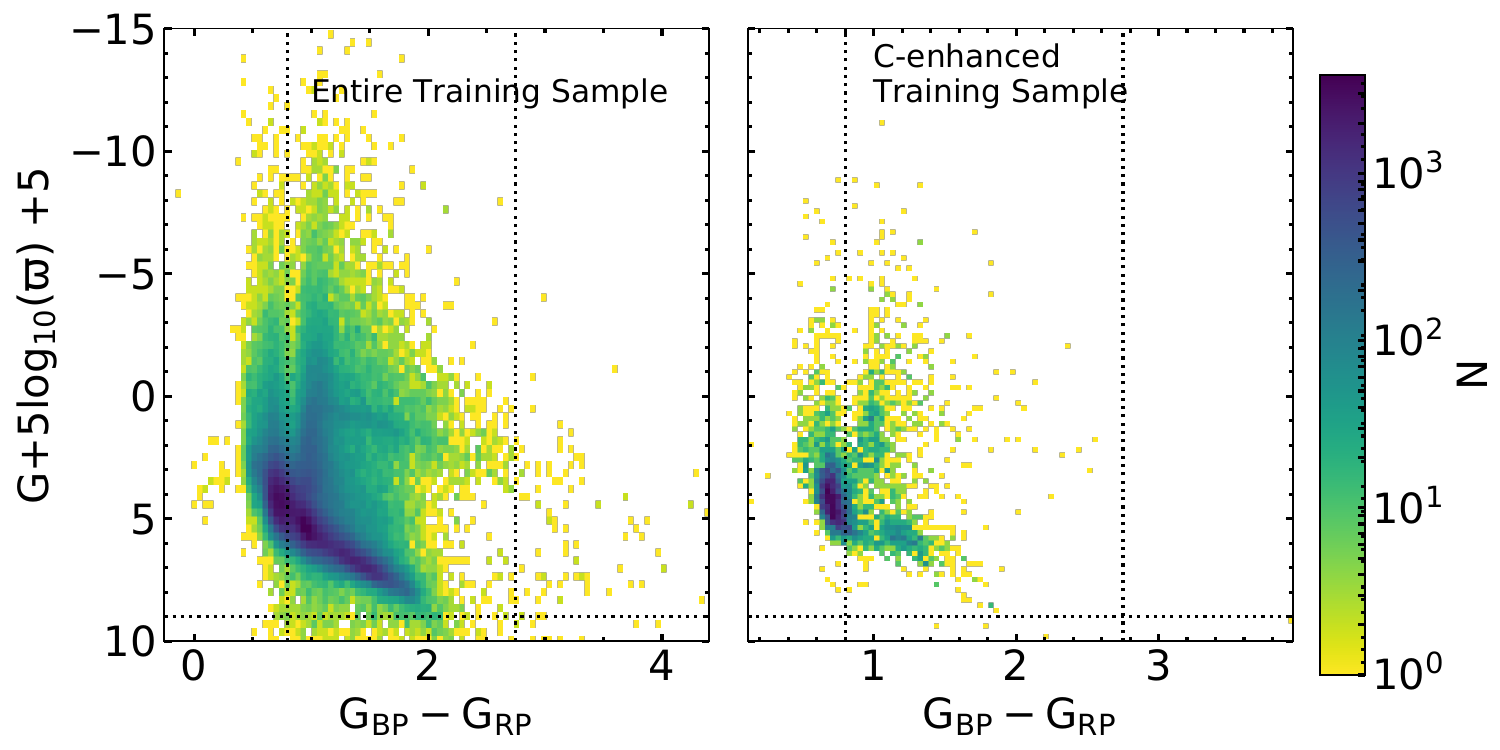}
    \caption{Color-magnitude diagram of our training sample from SDSS/SEGUE. The x-axis is the \gaia\ \bprp\ color;  the y-axis is the absolute $G$ magnitude, calculated from the distance modulus using the \gaia\ apparent $G$ magnitude and parallax. The left panel shows the entire training set, while the right panel only shows stars with [C/Fe] $>$ +0.7. The black dashed lines correspond to the color and magnitudes cuts made on our training/testing sample. The logarithmic color bar corresponds to the number of stars for each data point}
    \label{fig:train}
\end{figure*}

\begin{figure*}

    \includegraphics[width=\linewidth]{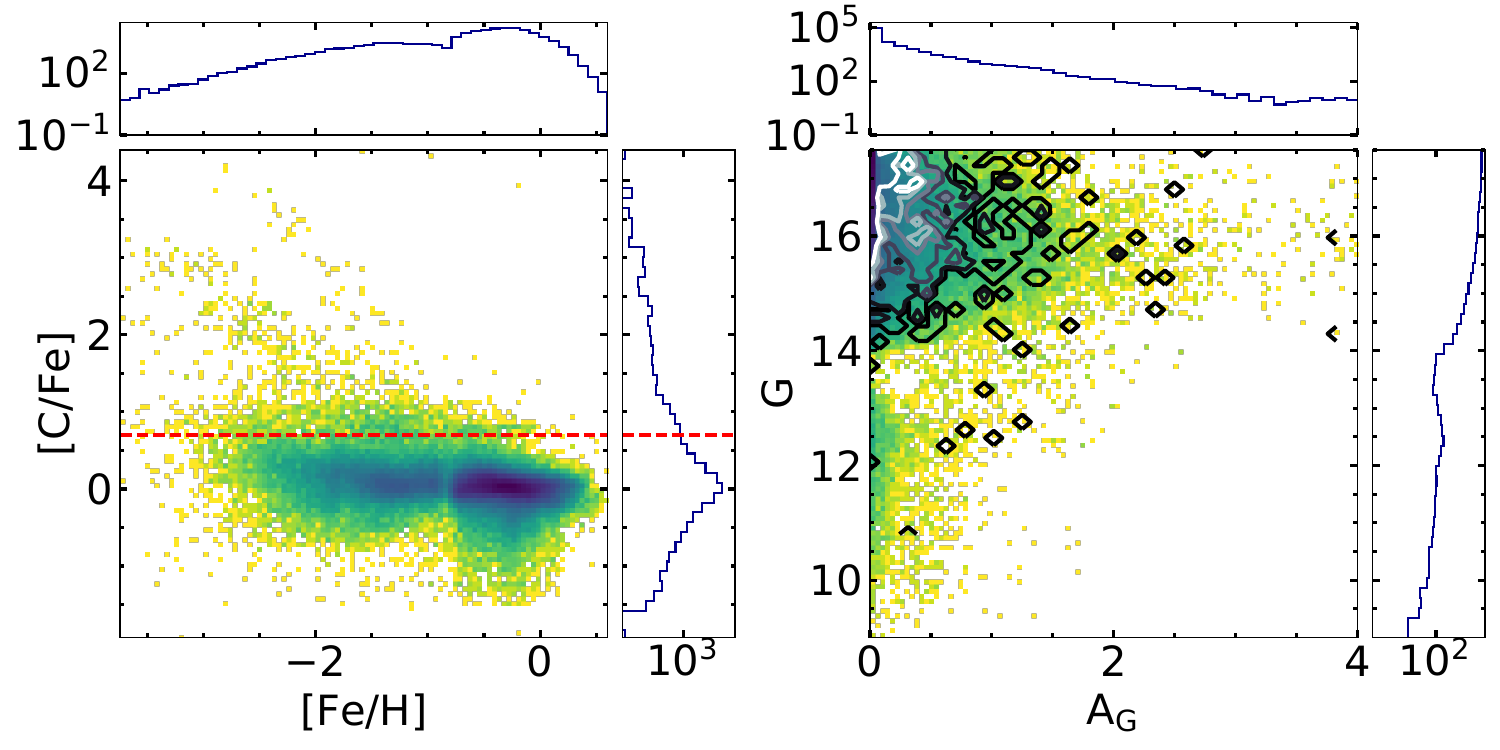}
    \caption{Relevant properties of the training/testing sample from the SDSS/SEGUE. The left panel shows the carbonicity ([C/Fe]), as a function of metallicity ([Fe/H]), with our definition of carbon-enhanced ([C/Fe] > +0.7) marked as a red dashed line. We also show the marginal histogram of each parameter on the corresponding axis. The right panel is a similar plot, but the axes are instead the extinction \citep[$\rm{A_G}$;][]{Andrae2022} and apparent \gaia\ $G$ magnitude. From inspection, the training/testing samples span a large range of parameters, similar to what we expect for the data we classify. In both plots the colors of the data points correspond to the logarithmic color bar shown in Fig.\ref{fig:train}. In the right panel, we also overlay the distribution of the CEMP stars in the training/testing sample as black/white contour lines. }
    \label{fig:train2}
\end{figure*}

In order to teach our model how to accurately detect carbon-enhanced stars, we require a sample of stars that are already classified. To acquire this, we employ the spectroscopic catalogs of parameters for stars from the SDSS survey, and its various extensions, including the Sloan Extension for Galactic Understanding and Exploration \citep[SEGUE;][]{Yanny2009,Lee2013,Rockosi2022} , which obtained over 500,000 low-resolution ($R$ = 2000) optical spectra. For simplicity, we refer to this collection of spectra as the SDSS/SEGUE sample.

The SEGUE Stellar Parameter Pipeline (SSPP;  \citealt{allende2008,lee2008a,lee2008b,lee2011,
smolinski2011,Lee2013}) has continued to be refined since its introduction. In the process, additional calibration stars with available high-resolution spectroscopic analyses have been used to improve estimates of the stellar parameters from the SDSS/SEGUE spectra.  The most recent version of the SSPP has been run through the stellar samples we employ. Note that, at this stage, the spectra include examples of objects originally targeted as QSO or candidate galaxy candidates, but that turned out to have spectra that were stellar in appearance. This is important, since numerous late-type CEMP stars turned out to be originally targeted as QSOs, based on their photometry (strong carbon absorption features can lead to colors that mimic quasars). 

After removal of duplicates (retaining the parameter estimates for the highest signal-to-noise ($S/N$) spectrum among repeated objects), the full set of spectra were then inspected visually (by Beers), for the identification of defective spectra that could perturb the stellar parameter estimates, the identification and rejection of white dwarfs, some of which were missed by the flags raised by the SSPP, as well as likely spectroscopic binaries, often comprising a white dwarf and a late-type star.  For the recognized spectroscopic binaries that do not include a white dwarf, the estimated stellar parameters are not necessarily compromised, as the SSPP parameters primarily consider features in the bluer portion of the spectrum.  Nevertheless, we conservatively dropped them from inclusion.  Care is taken in order to evaluate the best available estimate of [Fe/H], based on consideration of the various techniques available in the SSPP.  

Note that, for the purpose of the present application, we use the measured carbon abundances, without corrections for evolutionary effects (e.g., the first and second dredge ups, see \citealt{Placco2014}). This means that there are stars that appear to have [C/Fe] estimates below our adopted CEMP cutoff, but would appear above this cutoff once corrections are applied.  This is appropriate since, at this stage, we are primarily interested in identifying candidate CEMP stars in the \gaia\ DR3 sample, and this conservative choice ensures that we do not miss-classify stars due to uncertainties in the corrections.

After the culling procedure described above, we are left with 569,874 stars, of which 29,399 have [C/Fe] $>$ +0.7. We find that 233,604 of the original 569,874 stars have BP/RP spectra released in \gaia\ DR3, of which 9,094 ($\approx$4\%) have [C/Fe] $>$ +0.7. In preliminary tests, we found that our algorithm is incapable of detecting carbon enhancement in warmer stars, due to the weakness of the molecular carbon bands, so we apply a cut in the \bprp\ color. Specifically, we only include stars with \bprp\ $>$ 0.8, which roughly corresponds to \teff\ $\approx$6000\,K \citep{Andrae2018}. This cut is consistent with results from \citet{Witten2022}, who found they could achieve a carbon-abundance precision of $\approx$0.5 dex for stars with [Fe/H] $< -2$ only if they restrict their analysis to stars with \teff\ $<$ 6000\,K. This final trim leaves samples of 1,514 carbon-enhanced stars and 141,108 carbon-normal stars.

 We randomly select $\approx$ 30\% of these data as our testing sample, while the remaining $\approx$ 70\% is used for training. Figure \ref{fig:train} shows color-magnitude diagrams of our training sample. The left panel shows the \gaia\ DR3 \bprp\ color on the x-axis and the absolute $G$ magnitude calculated using the \gaia\ parallax and apparent $G$ magnitude on the y-axis for the entire training/testing sample.
The right panel shows the part of the sample that has [C/Fe] $>$ +0.7. To avoid extrapolation, we classify only stars that fall within this color-magnitude distribution (dotted black lines). However, note that our training sample is not uniformly distributed in this space, which may introduce a bias in our classification. We investigate this by evaluating the false positive and true positives rates  of the classification as a function of color and absolute $G$ magnitude (see Section \ref{sec:p&c}).

Figure \ref{fig:train2} shows the carbonicity as a function of metallicity (left panel), as well as the apparent $G$ magnitude as a function of extinction (right panel) for the training/testing sample. In the left panel, we also show our definition for carbon enhancement as a red dashed line at [C/Fe] = +0.7. It is interesting to note the appearance of two sequences with different slopes in the carbonicity as a function of [Fe/H] plane for stars with [C/Fe] $>$ +1. Most of the 1514 carbon-enhanced stars are metal-poor ([Fe/H] $\leq-1$). However, there are 139 carbon-enhanced stars with [Fe/H] $> -1$, corresponding to 9\% of the stars. We find that the classification algorithm identifies CEMP stars more accurately when these are included as positive cases in the training. We expect the metallicity distribution of carbon-enhanced stars in \gaia\ DR3 to be similar to our training/testing sample. Therefore, we call the stars positively classified by our algorithm as CEMP candidates, as we expect most of them to be metal-poor with only $\approx$9\% to be metal-rich.

In the right panel of Fig. \ref{fig:train2}, the extinction values are from the \gaia\ DR3 GSP-Phot pipeline \citep{Andrae2022}. Our training/testing sample includes data with $\rm{A_V}\approx$ 4. Given that we constrain the data we wish to classify to 0.8 < \bprp\ < 2.75, it is unlikely we would include any data with $\rm{A_V} >$ 4. Therefore, our training/testing sample should sufficiently teach our algorithm to distinguish highly extincted stars from carbon-enhanced stars. In addition, we can use the apparent $G$ magnitude distribution to investigate the signal-to-noise ratio (SNR) distribution of our testing/training sample. Given that the majority of the SDSS/SEGUE data is fainter than the apparent $G$ magnitude cut for the BP/RP spectra released in DR3 ($G\leq$ 17.5), we find the apparent $G$  magnitude distribution of the SDSS/SEGUE data peaks at $G\approx$ 17.5. Therefore, the $G$ magnitude distribution of our training and testing data matches our expectations for the BP/RP spectra that we classify in that it peaks at the faintest magnitudes, but also includes stars as bright as $G \approx$ 10.

\subsection{Gaia BP/RP Spectra}
\label{sec:bprp}
The \textit{Gaia} BP/RP spectra are a unique data set, not only due to their wide wavelength coverage, very low-resolution ($R \approx$ 50), and an unprecedented number of stars, but also because the spectra have been released as a linear combination of basis functions, specifically as Hermite function coefficients \citep{Carrasco2021}. This was done because of the complexity of the \gaia\ instrument, which has two wide fields of view and 14 detectors. To create the calibrated mean spectra, multiple epochs of observations with different instrumental conditions needed to be combined. In this work, we use the coefficients as the input data for our model rather than convert them to sampled spectra, which results in some information loss \citep{Carrasco2021}. 

In total, the BP and RP spectra comprise 55 coefficients each, but also come with a recommended truncation. This is possible because the coefficients have been rotated to an optimized basis so that the bulk of the spectral information is contained in the first few coefficients. A truncation is then recommended based on the magnitude of the coefficients compared to their corresponding uncertainties. For more details see \citet{Carrasco2021}. As \texttt{XGBoost} requires the input data to be vectors of the same length, we apply the largest recommended truncation to avoid losing potentially useful information. The largest recommended truncation is 55 for both BP and RP spectra (i.e., all coefficients are relevant). Therefore, we do not truncate the coefficients. Because we do not want to include apparent magnitude information, we normalize the coefficients by the first BP coefficient. Furthermore, since the coefficient values can span many orders of magnitude, we  also divide the spectra by the mean normalized spectrum of the training sample. This decreases the orders of magnitudes spanned by the coefficients, which makes it easier for XGBoost to find the 
carbon-enhancement signal in the data.

The \texttt{XGBoost} algorithm cannot reliably extrapolate. Therefore, we ensure to only classify stars that are similar to our training sample. Specifically, we constrain  our sample using the absolute $G$ magnitude and \bprp\ color. As shown in the right panel of Figure \ref{fig:train} by the black dotted lines, we only classify stars that satisfy the following criteria: 
\begin{enumerate}
    \item $0.8 < G_{BP}-G_{RP} < 2.75$
    \item $ G + 5\rm{log_{10}(\varpi)} + 5 < 7.0$
\end{enumerate}
Although our training sample includes stars outside of this range, we chose to restrict to these values where most of the sample resides. Note that we do not place a lower limit on the absolute $G$ magnitude, since stars with very low absolute $G$ magnitudes likely have small, uncertain parallaxes that cause an over-estimation in the brightness, and therefore underestimate the absolute $G$ magnitude. As we do not want to introduce a selection bias by removing stars with uncertain parallaxes, we chose to include them. In Section \ref{sec:p&c}, we investigate how the false positive rate and completeness (i.e., the true positive rate) of our classification behaves at the edges of this region where the training sample is less dense. In total, we find 182,815,672 BP/RP spectra in \gaia\ DR3 that are within our color and magnitude cuts.

\section{\texttt{XGBoost}} \label{sec:xgb}

\begin{figure*}

    \includegraphics[width=\linewidth]{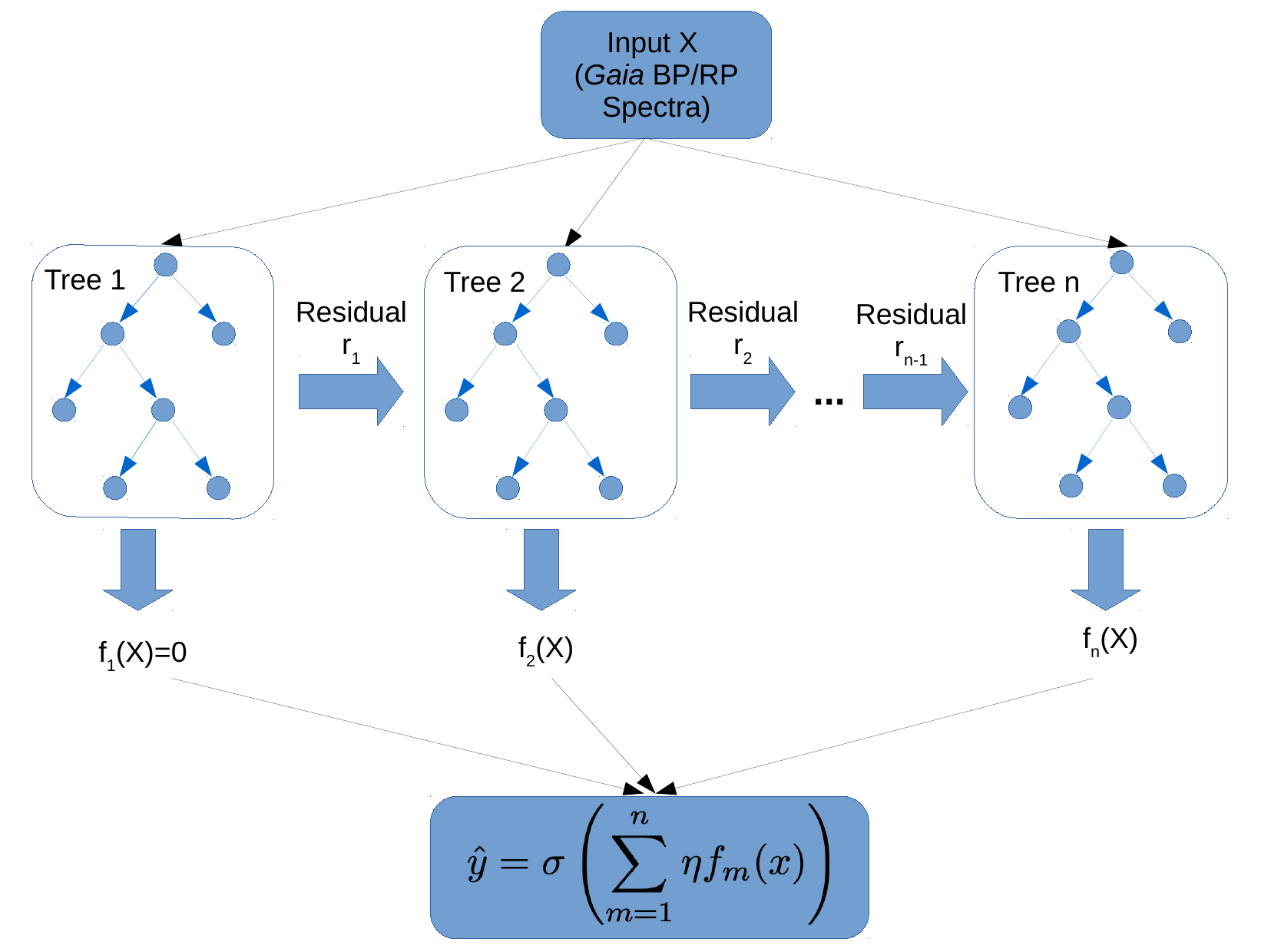}
    \caption{General architecture of \texttt{XGBoost} for classification. In short, \texttt{XGBoost} iteratively creates trees to fit the residuals from the prediction of the previous tree. The first tree provides a prediction of zero for each spectrum. For the final output, the predictions of each tree are summed after being multiplied by the learning rate, $\eta$. This value is then input into the sigmoid function, $\sigma(x) = 1/(1+e^{-x})$, to calculate the final probability, $\hat y$. If $\hat y>$ 0.5, the star is classified as carbon enhanced.}
    \label{fig:xgb}
\end{figure*}

In order to detect carbon-enhanced stars across a wide range of stellar parameters and reddenings, we require a flexible model. We chose to use \texttt{XGBoost}, which is powerful but still easy to interpret. Furthermore, \texttt{XGBoost} is optimized for efficiency, allowing fast training and inference. \texttt{XGBoost} is quickly becoming a popular machine learning algorithm in astronomy, with applications in a large variety of sub-fields \citep[e.g.,][]{Hayden2020,Valle2021,Li2021,He2022,Pham2022}

Figure \ref{fig:xgb} shows the general architecture of \texttt{XGBoost} used for classification. For a detailed description of the algorithm see \citet{Chen2016}. In short, \texttt{XGBoost} sequentially builds decision trees to fit the residuals from the previous tree. \texttt{XGBoost} continues to train trees until it reaches the maximum number of trees set by the user or the residuals stop consistently shrinking. The results from each tree are then summed together, weighted by the learning rate, $\eta$. This value is then plugged into the sigmoid function, $\sigma (x)= 1/(1+e^{-x})$, to calculate the probability that the star is carbon enhanced ($\hat y$). We provide these probability values so that the reader can choose their own sample depending on the completeness and contamination rate required for their science. In this work, we choose to classify a star as carbon enhanced if its probability is > 50\%.

\texttt{XGBoost} does not allow for the direct inclusion of uncertainties for each input, but it is able to learn how to distinguish noise from signal sufficiently if the noise distribution of the training sample is representative of the data to which the model will be applied. Given that the apparent $G$ magnitude distribution of our training sample is similar to what we expect for the BP/RP spectra we classify (see Figure \ref{fig:train2}), we conclude that the noise distribution of the training sample is representative and sufficient to train the \texttt{XGBoost} model. 

To train the \texttt{XGBoost} algorithm, a number of hyperparameters need to be set. We can set the maximum number of trees, the learning rate ($\eta$), the percentage of the training sample and the percentage of input coefficients to use for each tree, as well as the maximum depth of each tree. We can also set limits on the purity of a sample for a given leaf to prevent overfitting. To explore the parameter space and find the optimal set of hyperparameters, we use \texttt{RandomSearchCV} from scikit-learn \citep{scikit-learn}.

\section{Contamination and Completeness} \label{sec:p&c}

\begin{figure}

    \includegraphics[width=\linewidth]{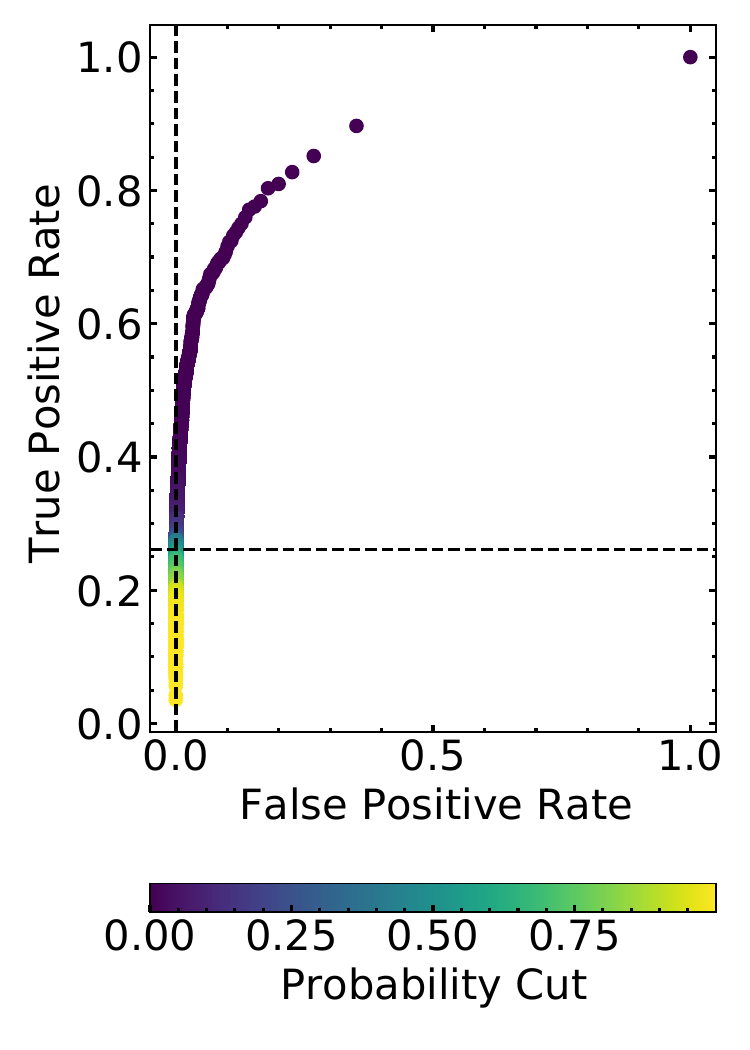}
    \caption{The receiver operating characteristic (ROC) describing the false positive (false positives divided by the sum of the true negatives and false positives) and true positive rates (true positives divided by the sum of the false negatives and true positives) of our classification for the testing sample. We calculate this curve by assuming different probability cuts for our classification, which are shown by the color of the points. We mark the point where the probability cut is > 50\% with black dashed vertical and horizontal lines. Given that only $\approx$ 1\% of our training sample is carbon enhanced, the classification is very unbalanced. Therefore, the false positive rate is very small, even though the contamination rate (false positives divided by the sum of the true positives and false positives) is $\approx$ 12\%. The true positive percentage or completeness is 26\%.}
    \label{fig:roc}
\end{figure}

\begin{figure}

    \includegraphics[width=\linewidth]{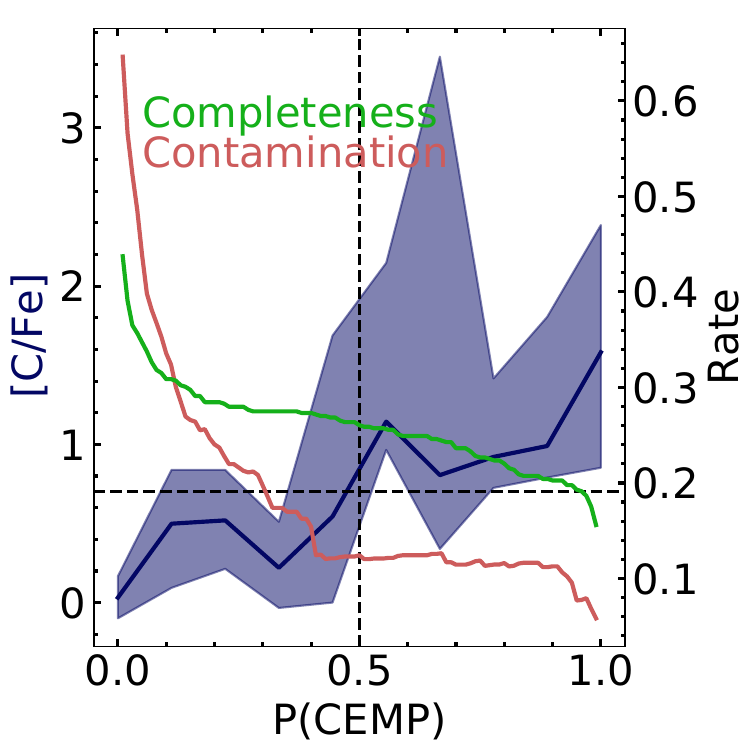}
    \caption{ The median carbon abundance of the testing sample, as a function of the assigned probability of carbon enhancement, is shown by the dark blue line, with the 1$\sigma$ percentiles shown in the blue shaded region. We also show the contamination rate as a red line, and the completeness as a green line, with the scaling shown on the right y-axis. The vertical dashed line corresponds to $p$(CEMP) = 0.5 and the horizontal dashed line corresponds to [C/Fe] = +0.7, above which is our definition of a carbon-enhanced star. It is clear that the assigned $p$(CEMP) is strongly correlated to carbon abundance. Furthermore, we find that the algorithm learns our definition of carbon-enhanced is [C/Fe] > +0.7, in that the median [C/Fe] for $p$(CEMP) $\approx$ 0.5 is $\approx$ +0.7.} 
    \label{fig:pcut}
\end{figure}

\begin{figure*}

    \includegraphics[width=\linewidth]{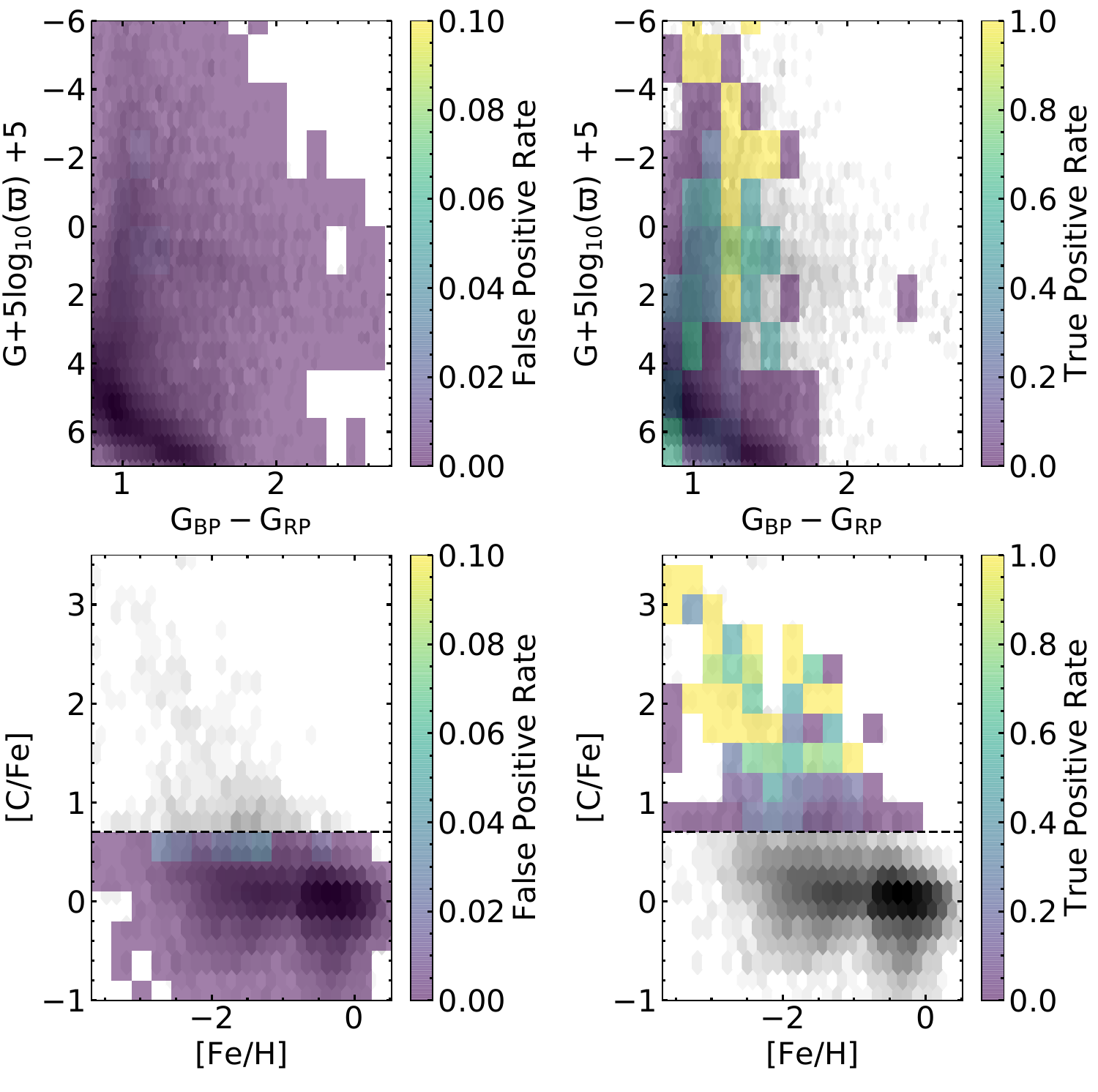}
    \caption{ The false positive and true positive rates, as functions of absolute $G$ magnitude, \bprp\ color, [C/Fe], and [Fe/H]. We show the distribution of stellar parameters for the testing sample in greyscale, with darker areas corresponding to more stars.  In general, the model tends to struggle most with red dwarf stars and blue giant stars. Furthermore, the false positives tend to have +0.5 $\leq$ [C/Fe] $\leq$ +0.7, while the false negatives tend to have +0.7 $\leq$[C/Fe] $\leq$ +1.0.  Therefore, our model likely only can interpret the carbon abundance to $\approx$ 0.5 dex for some stars. }
    \label{fig:kc}
\end{figure*}

\begin{figure}

    \includegraphics[width=.99\columnwidth]{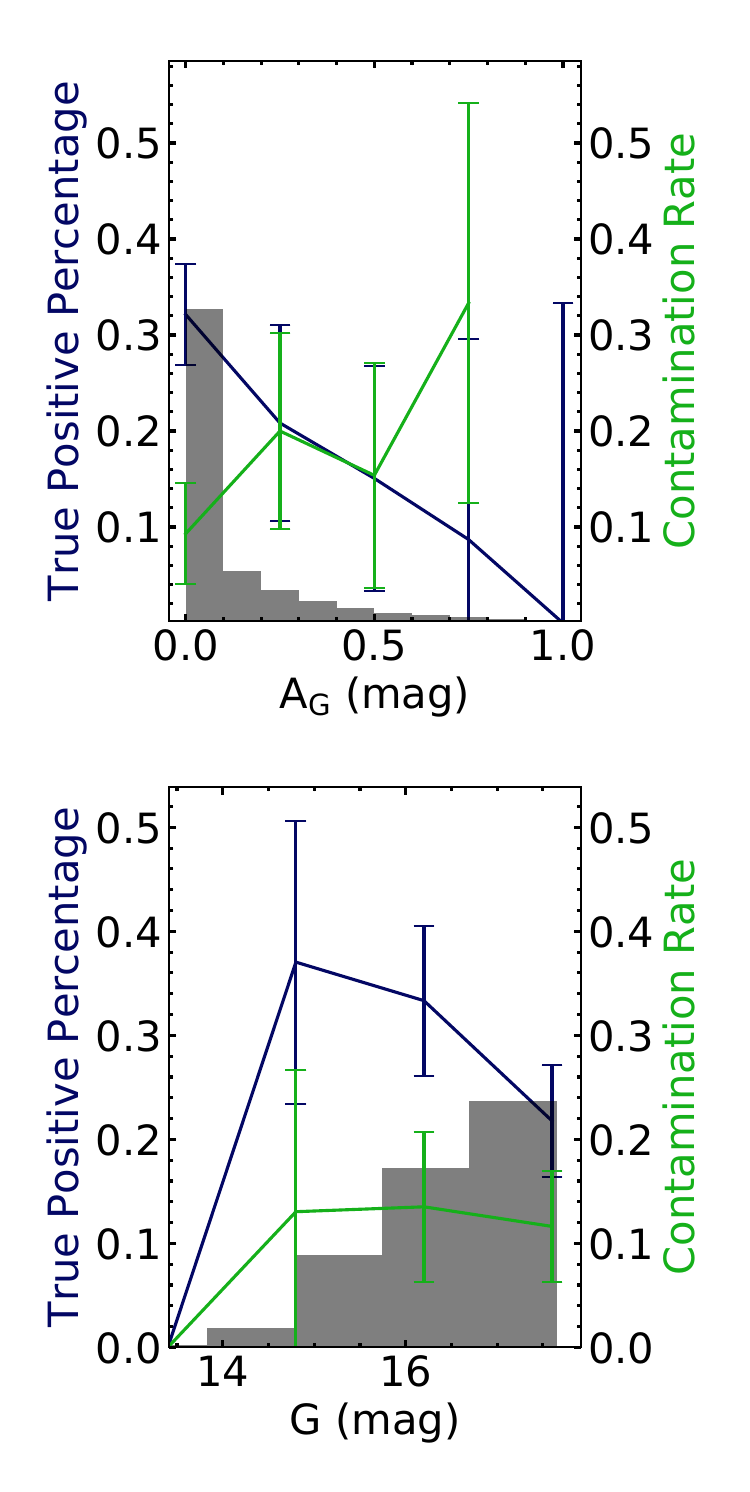}
    \caption{The contamination rate and true positive percentage, as a function of $G$ magnitude and extinction ($\rm{A_G}$). The grey bins show the arbitrarily scaled number density of stars in our training/testing sample. For each panel, the scale for the true positive percentage (dark blue) is on the left y-axis; the right y-axis shows the scale for the contamination rate (green). The error bars are 1/$\sqrt{N}$, where $N$ is the number of true positives. As expected, we find that our classification improves for bright stars, which likely have high signal-to-noise spectra, and for stars at low extinction.}
    \label{fig:ac}
\end{figure}

In order to estimate the contamination and completeness (i.e., true positive rate) of our sample of newly identified CEMP candidates, we use our testing sample (described in Section \ref{sec:train}), where we already know the observed carbon abundances. Specifically, we define contamination as the number of false positives divided by the sum of the true positives and false positives. In other words, the contamination estimates the rate of carbon-normal stars in the sample we classify as CEMP. On the other hand, we define completeness as the number of true positives divided by the sum of the true positives and false negatives. Therefore, the completeness estimates the fraction of true CEMP stars that we expect to detect. There may be non-stellar objects that contaminate our sample (e.g., quasars), but we mitigate these effects by constraining the full set of spectra that we classify by the absolute $G$ magnitude and $\rm{G_{BP}-G_{RP}}$ color of the training/testing sample (see Section \ref{sec:bprp}). Furthermore, stars hotter than our training/testing sample might mistakenly be included if high levels of extinction make them appear sufficiently red. However, we expect that this situation is rare, given that our training sample is representative of the full distribution in \teff\ and extinction (see Figure \ref{fig:train} and \ref{fig:train2}).  In addition, we have stellar parameters and carbon abundances for the testing sample, which allows us to study how the contamination and completeness behave as a function of various parameters, including observational effects. 

 Figure \ref{fig:roc} shows the receiver operating characteristic (ROC) curve, which describes the false positive rate (false positives divided by the sum of the true negatives and false positives) and true positive rate (true positives divided by the sum of the false negatives and true positives) of the classification as a function of the probability cutoff assumed. As expected, the true positive percentage and false positive rate increase as the probability cut is decreased. The black dashed vertical and horizontal lines mark the true positive and false positive rates for a probability cut of > 50\%, which we use to define the CEMP candidate sample in this work. Specifically, for a probability cut of > 50\% we find a false positive rate of 0.04\% and a true positive rate of 26\%. The false positive rate is especially low because our classification is unbalanced, with only $\approx$ 1\% of our sample being a positive case (i.e., carbon enhanced). On the other hand, the contamination rate (false positives divided by the sum of the false positives and the true positives) is $\approx$ 12\%. In the final catalog, we provide the probability values for each star classified so that the reader can choose a probability cut best suited for their science case. 

Figure \ref{fig:pcut} shows the median carbon abundance of the testing sample, as a function of the assigned probability of being carbon enhanced, as a dark blue line, with the 1$\sigma$ percentiles as the shaded region. We also show the completeness (green line) and contamination rate (red line) for $p$(CEMP) > $x$ with the scaling on the right y-axis. The vertical black dashed lines gives the $p$(CEMP) that above which a star is classified as CEMP ($p$(CEMP) = 0.5). Therefore, where this line intersects the completeness (26\%) and contamination rates (12\%) gives those properties for our final sample. The horizontal dashed line indicates the [C/Fe] above which we define a star to be carbon-enhanced ([C/Fe] = +0.7). The median [C/Fe] becomes larger than this at $\approx$ $p$(CEMP) = 0.5, indicating that our algorithm learns to assign a $p$(CEMP) > 0.5 for stars with [C/Fe] > +0.7.

Figure \ref{fig:kc} shows 2D maps of the false positive (left) and true positive rates (right), as a function of absolute $G$ magnitude, and \bprp\ color (top), as well as for [C/Fe] and [Fe/H] (bottom). In these plots, we also show the underlying density distribution of stars as grey hexagonal bins. The left panels show the false positive rate, which is calculated by taking the number of false positives divided by the number of true carbon-normal stars in that bin. The right panels show the true positive rate, which is calculated by dividing the number of true positives (i.e., correctly identified CEMP stars) by the total number of true CEMP stars in that bin.

In the top left panel of Figure \ref{fig:kc}, we do not see a trend in the false positive rate with the \bprp\ color or absolute $G$ magnitude. In the top right panel, we see that at the true positive rate is lowest for dwarf stars and blue giant stars. We also find that the algorithm does not detect the most extincted, reddest giant stars in our testing sample. 

In the bottom row of Figure \ref{fig:kc}, we see a slight trend in the false positive and true positive rates with the [C/Fe] abundance. Specifically, we see the false positive rate is higher for stars with +0.5 < [C/Fe] < +0.7, and the true positive rate is lowest for stars with +0.7 < [C/Fe] < +1.0. This indicates that our classification is most inaccurate for stars with [C/Fe] $\approx$ +0.7, which is to be expected given that it is unlikely we could measure a [C/Fe] abundance from these very low-resolution ($R \approx$50) spectra that is more precise than $\approx$0.5 dex.

Figure \ref{fig:ac} shows the true positive percentage and contamination rate, as a function of extinction (top panel) and apparent $G$ magnitude (bottom panel). The grey histogram shows the arbitrarily scaled underlying distribution of the testing sample for these parameters. The green line shows the contamination rate, with the scaling provided by the right y-axis, while the dark blue lines show the true positive percentage, with corresponding scaling on the left y-axis.

As expected, we find that the classification performs better at low extinction. Specifically, we see that the true positive percentage decreases from $\approx$ 30\% at $\rm{A_G}$=0 to $\approx$0\% at $\rm{A_G}$=1. We also see that the contamination rate increases from $\approx$10\% to $\approx$ 30\% over the same range. Of our final sample of CEMP candidates, 77\% has $\rm{A_G}$ < 
0.5. Furthermore, we find that the contamination rate also increases with fainter $G$ magnitude, as expected, since the $G$ magnitudes are directly related to the signal-to-noise for these spectra. We find that the false positive percentage increases from $\approx$ 5\% at $G$ magnitude of 
$\approx$ 13 to $\approx$ 12\% at $G$ =17.5. The true positive percentage is lowest for bright stars. This is likely because there are few bright CEMP stars in our training sample (see Figure \ref{fig:train2}), which may introduce a bias against detecting bright CEMP stars in the \gaia\ data. We also see that the true positive percentage decreases at $G$ > 15. This is likely due to lower signal-to-noise for these data.

To further validate our sample, we also compare to large spectroscopic surveys. However, as CEMP stars are (relatively) rare, large spectroscopic surveys generally do not correctly account for CEMP stars in their analysis pipelines. For example, APOGEE's ASPCAP pipeline does not account for stars with [C/Fe] > +1 \citep{Abdurrouf2022}. When cross-matched against our SDSS training/testing sample, we find 1,765 stars in common with APOGEE DR17. Of these, only 26 are identified as CEMP in the SDSS/SEGUE catalog. However, the ASPCAP pipeline assigns only 1 of these stars an abundance of [C/Fe] > +0.7. This star has [C/Fe] = +2.28 from SDSS/SEGUE, while ASPCAP measures [C/Fe] = +0.89. From this comparison, we expect ASPCAP to report [C/Fe] < +0.7 for as much as 96\% of our final CEMP catalog. Of the stars we classify as CEMP, we find that 442 stars have a match in APOGEE DR17 without flags on the ASPCAP derived [C/Fe]. Of these, 425 have [C/Fe] < +0.7 as measured by ASPCAP, corresponding to 95\%. As we found that ASPCAP measured [C/Fe] < +0.7 for 96\% of our CEMP training sample, it is not unexpected that we find a high percentage of our final sample has [C/Fe] < +0.7 from ASPCAP as well. 

The metallicities measured by APOGEE of the stars that have $p$(CEMP)$>$ 0.5 in our catalog are generally metal poor, with mean [Fe/H] = --0.95 (compared to mean [Fe/H] = --0.25 for stars with $p$(CEMP)$<$ 0.5) and minimum [Fe/H] = --2.41. However, we note that only 17 of these stars have [C/Fe] $>$ +0.7 in APOGEE. Of these 17, APOGEE measures an average [Fe/H] = --1.39, with minimum [Fe/H]= --2.23. 

To validate the completeness of our sample, we cross-match against a database of CEMP stars confirmed from high-resolution spectra \citep{zepeda2023}. Of the 382 stars from \citet{zepeda2023} that have observed [C/Fe] $>$ +0.7 (no evolutionary correction), 208 of them have BP/RP spectra within our color and absolute magnitude cuts. We correctly classify 143 of those as CEMP stars, which gives a completeness rate of 69\%. This is significantly higher than that estimated with our testing sample (26\%). Therefore, it is possible that our true completeness percentage is higher, and 26\% is a conservative estimate. In general, we find that the completeness relative to \citet{zepeda2023} behaves as expected from Figures \ref{fig:kc} and \ref{fig:ac}. Specifically, the completeness decreases with decreasing [C/Fe], increasing \teff, and fainter apparent G magnitudes. We also find that we tend to miss Group II stars in \citet{zepeda2023} with lower [Fe/H], because they have lower absolute carbon abundance, $A$(C), and therefore weaker absorption features. 

\section{Model Interpretation} \label{sec:interp}

\begin{figure*}

    \includegraphics[width=\linewidth]{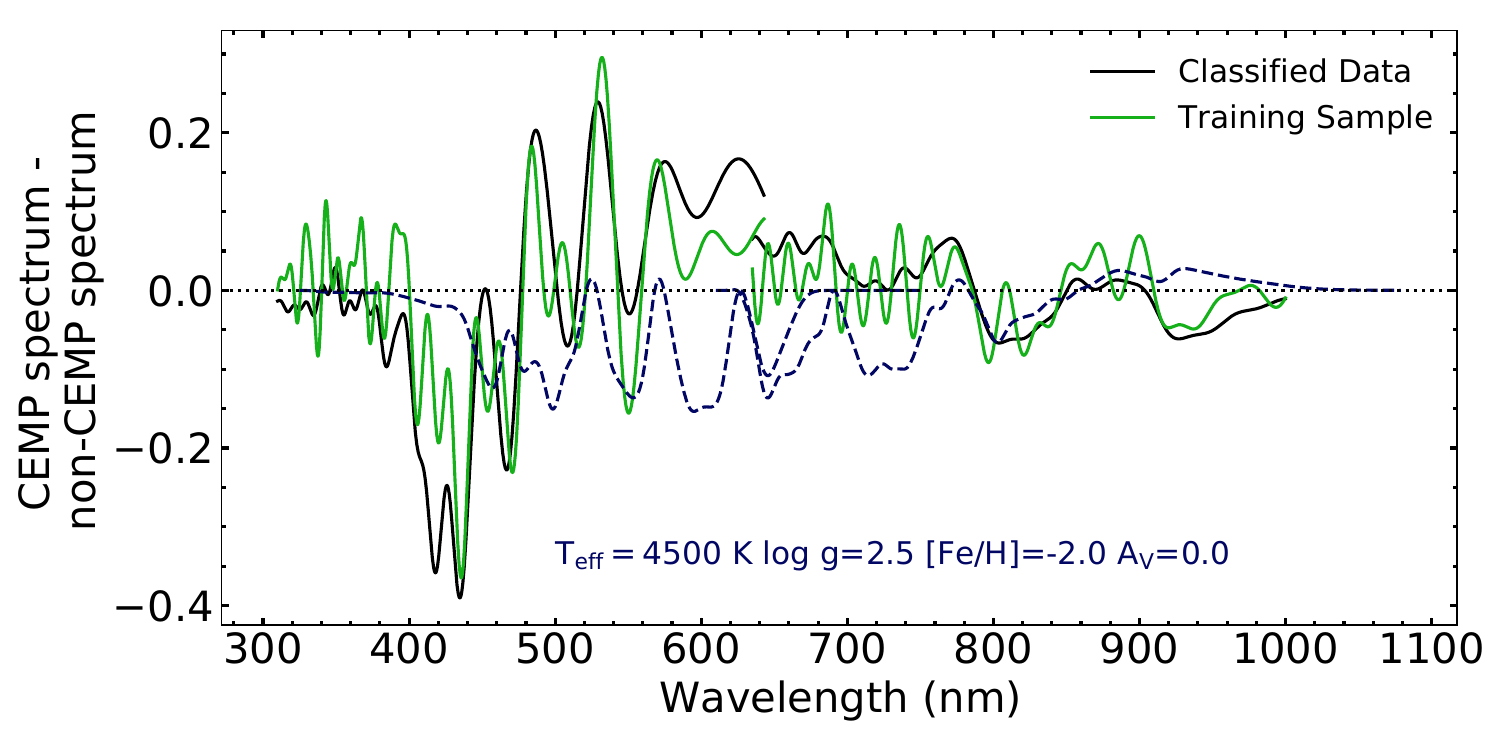}
    \caption{The average difference between CEMP spectra and carbon-normal spectra for a subset of the training sample (green) and newly classified data (black). We also include the difference between synthetic spectra (dark blue dashed line) with [C/Fe] = +1 and [C/Fe] = +0.5, assuming \teff\ = 4500\,K, \logg\ = 2.5, and [Fe/H] = -2.0 , as described in Figure \ref{fig:spec}. The green and black lines are calculated by subtracting the average spectrum of carbon-normal stars from CEMP stars for stars in a narrow range of color (1.20 $<$ \bprp\ $<$ 1.25) and absolute $G$ magnitudes (0 $<$ G+5$\rm{log_{10}}(\varpi)+5 <$ 3) which corresponds to a small dense region of the red giant branch.  We do this both for the training sample and the newly classified data in order to ensure that the classification has worked, and that the newly classified CEMP stars have the expected carbon features. Given that the difference between the CEMP stars and the carbon-normal stars is quite similar for both the training and newly classified samples, and that the features generally match expectations from synthetic spectra, we conclude that our classification has correctly selected candidate CEMP stars.} 
    \label{fig:spec_diff}
\end{figure*}

\begin{figure*}

    \includegraphics[width=\linewidth]{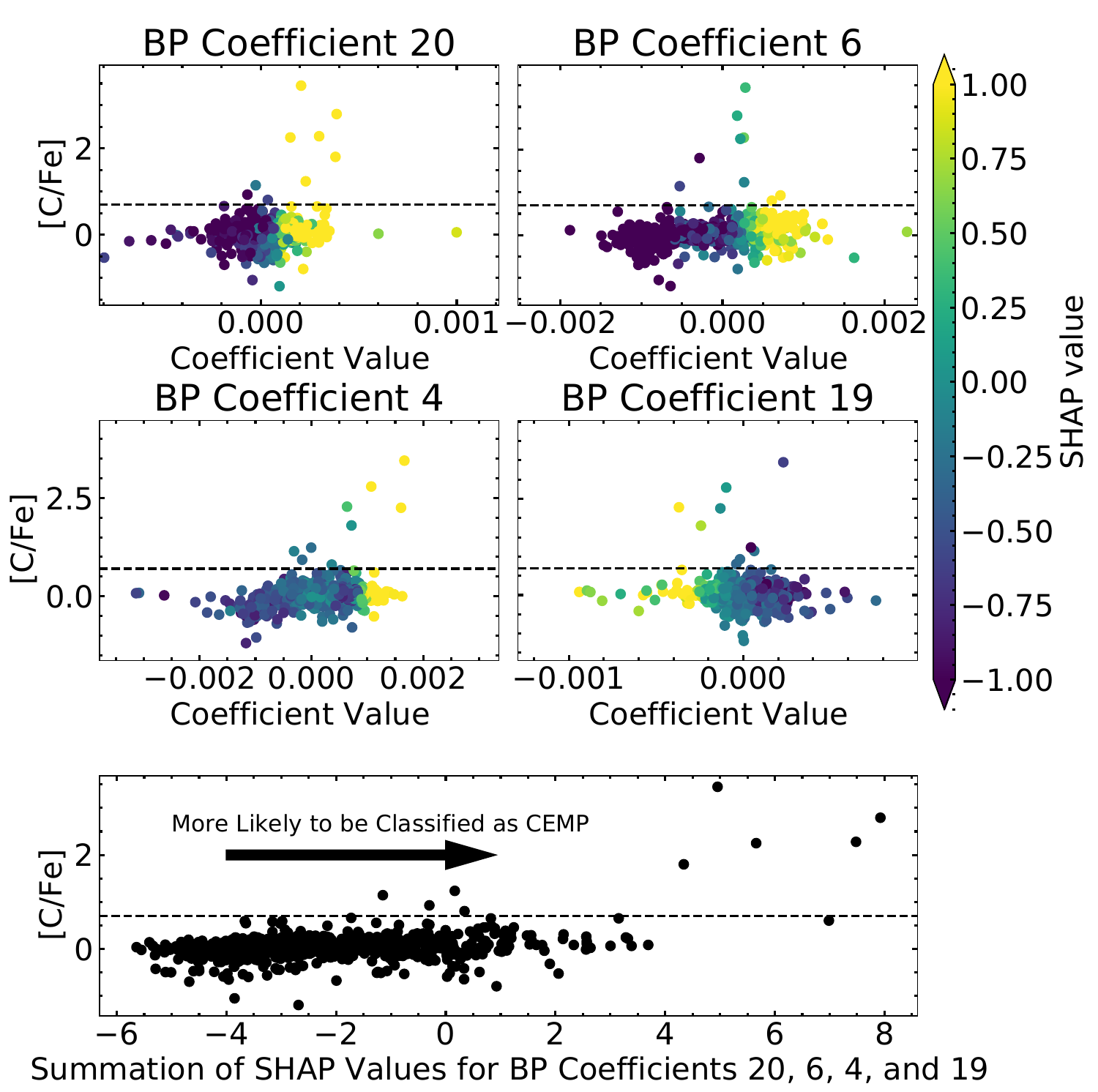}
    \caption{ The relations between [C/Fe], the spectral coefficient value, and the SHAP value for the four spectral coefficients with the largest SHAP values. We show a horizontal black dashed line at [C/Fe] = +0.7. The bottom panel shows [C/Fe] as a function of the summation of the SHAP values of the four coefficients. Each point corresponds to an individual star in our testing sample.  The SHAP values give the feature importance, in that large positive SHAP values indicate a large increase in the assigned probability of the star being carbon enhanced due to a given coefficient's value. Therefore, if the model uses the carbon information in the coefficients, we expect that stars with [C/Fe] > +0.7 to have high positive SHAP values.}
    \label{fig:shap}
\end{figure*}

As we have used a data-driven method to classify stars in this work, we want to ensure that the final model matches our physical intuition. It is also important to ensure that the model is not using any confounding variables (e.g., extinction or metallicity) that could be correlated with carbon enhancement but are not a direct measurement of the true carbon abundance. Typically, model interpretation is done by determining the importance of each input feature for the final inference and comparing it to expectations from physical models. Here, we investigate the importance of four key Hermite coefficients and how they relate to true carbon abundance. Furthermore, we compare BP/RP spectra of CEMP stars from the training sample to newly classified stars in order to ensure they match expectations.

Figure \ref{fig:spec_diff} shows the average difference between CEMP spectra and carbon-normal stars for a subset of the training sample (green) and the newly classified data (black). We also show the difference between synthetic spectra (dark blue dashed line) with [C/Fe] = +1 and [C/Fe] = +0.5, assuming \teff\ = 4500\,K,  \logg\ = 2.5, and [Fe/H] = --2. This is the same line as shown in the top right panel of Figure \ref{fig:spec}. For the observed spectra, we chose to calculate the averages over a narrow color and absolute $G$ magnitude range in order to isolate the effect of carbon from the effects of \teff, \logg, and extinction on the spectra. Specifically, we use stars with 1.20 $<$ \bprp\ $<$ 1.25 and 0 $< G+5\rm{log_{10}}(\varpi) +5 < $ 3, which is a narrow dense region of the red giant branch (RGB). For our training sample, this corresponds to a \teff\ range of 4500\,K $<$ \teff $<$ 5500\,K. We calculate the average CEMP spectrum and carbon-normal spectrum in this region for both the training sample and the newly classified sample, in order to ensure that they look similar, which indicates that the classification algorithm successfully identified candidate CEMP stars. We also compare this to expectations from synthetic spectra. We find that the differences between CEMP and carbon-normal stars for the newly classified data matches the differences found in the training sample. Furthermore, the features seen in the subtracted spectra roughly match with expectations from mock spectra. The major discrepancies are likely due to imperfect modeling and different normalization procedures. Specifically, the observed spectra are normalized in the \gaia\ BP/RP coefficient space following the procedure described in Section \ref{sec:bprp}, while the synthetic spectrum is normalized by the maximum flux. Therefore, the scaling of the features between the observed and model spectra should not be compared. In general, the model and observed spectra agree in the location of local minima and maxima, indicating agreement in the location of key absorption features. However, blue-ward of $\approx$480 nm, there is poor agreement between the model and observed spectra. This is likely due to a shortcoming in the model-spectra resolution. The resolution of the BP spectra increases in the bluer regions to $R\approx$100, but the mock model spectra assumes $R$=50 for the entire wavelength region.    

We measure the impact of each Hermite basis coefficient on the classification using SHAP (Shapley Additive exPlanations) values \citep{shapley_lundberg2017unified}. The SHAP values allow us to explore the importance of individual features, as a function of various parameters, since each star's coefficients are assigned individual values. The SHAP values are defined so that their summation plus the average predicted value ($\phi_0$) is equal to the predicted value ($\hat y$). Explicitly, 
\begin{equation}
    \hat y = \phi_0 + \sum_{i=1}^M \phi_i,
\end{equation}
where $\phi_0$ is the average value of $\hat y$ for all of the spectra, $\phi_i$ is the SHAP value for coefficient $i$, and $M$ is the total number of coefficients per spectrum. Therefore, each SHAP value directly measures the impact of the selected coefficient on the inference of $\hat y$ for a given star. We refer the interested reader to \citet{shapley_lundberg2017unified} for further details on the calculation of SHAP values.

We calculate the SHAP values for all of the spectra in our testing sample. Figure \ref{fig:shap} shows how the SHAP values relate to the carbon abundance and coefficient value for four BP coefficients. These coefficients were selected because they have the largest SHAP values of all the BP and RP coefficients. Specifically, we have the [C/Fe] on the y-axis, and the coefficient value on the x-axis. The points are colored by the SHAP values. The bottom panel shows [C/Fe] as a function of the summation of the four SHAP values. As large positive SHAP values indicate that the given coefficient increased the probability that the star is carbon enhanced, it is expected that stars with [C/Fe] > +0.7 should have higher SHAP values than carbon-normal stars.  We find that the \texttt{XGBoost} model is able to pick up this sensitivity, in that the SHAP values for these coefficients are  generally large and positive for carbon-enhanced stars. Although there are some coefficients where the SHAP value is negative for a carbon-enhanced star, these stars may still be classified as carbon enhanced based on the value of other coefficients. Similarly, there are many carbon-normal stars that have individual positive SHAP values, but the combined effect of all of the other coefficients effectively decreases the probability of the star being carbon enhanced (see the bottom panel), so that the false positive rate is not exceedingly high. Given the positive correlation between [C/Fe] and the SHAP value, it is likely that \texttt{XGBoost} model is using the carbon information in the spectra to determine whether a given star is carbon enhanced rather than using another confounding variable. 

\section{Properties of the CEMP Candidate Sample} \label{sec:sample}
\begin{figure}

    \includegraphics[width=\linewidth]{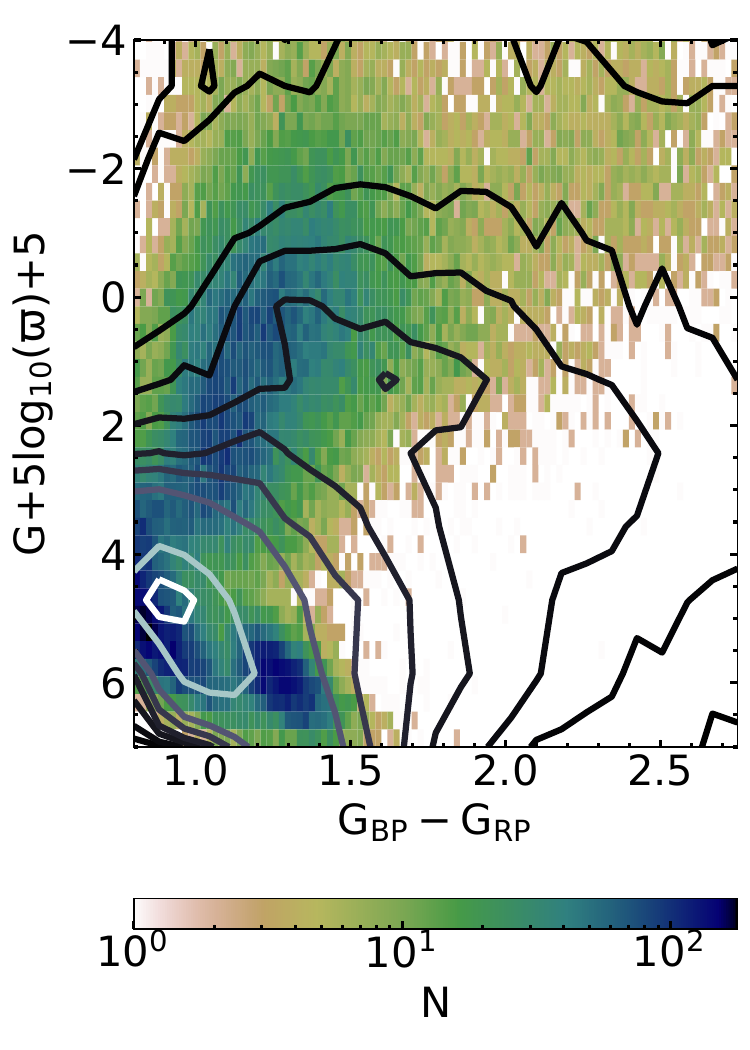}
    \caption{ Color-magnitude diagram of the final sample of candidate carbon-enhanced stars (dark blue/beige gradient), compared to a random sample of stars classified as not carbon-enhanced in white (high density) to black (low density) gradient contour lines. In general, the color-magnitude distribution of our CEMP candidate sample is similar to the carbon-normal sample, in that the majority of stars are bright dwarf/turn-off stars. On the giant branch, however, the CEMP candidate sample tends to be bluer, while the carbon-normal sample shows a clear red clump at absolute $G \approx$ 1. This indicates that the CEMP candidate sample is more metal-poor than the carbon-normal sample, as expected. }
    \label{fig:cmd_f}
\end{figure}

\begin{figure}

    \includegraphics[width=\linewidth]{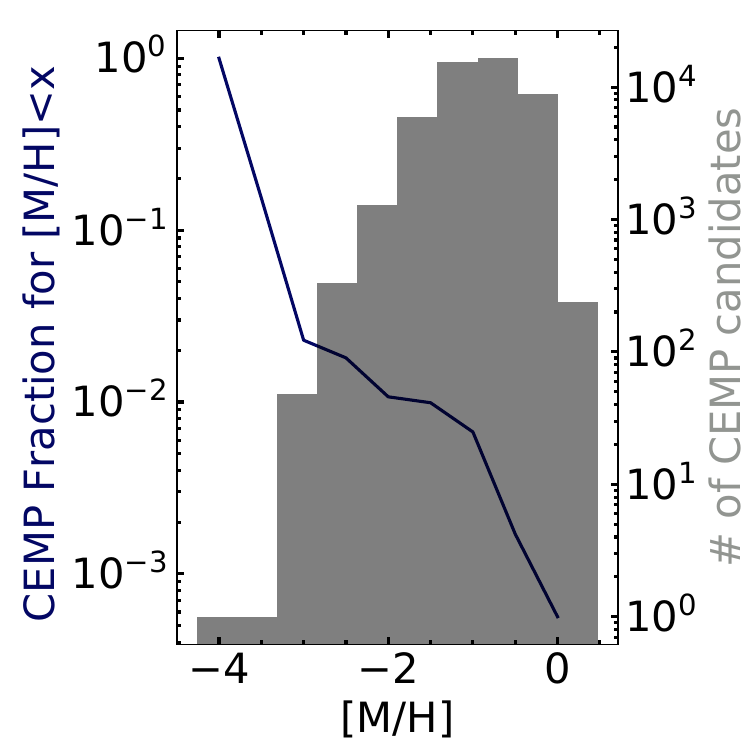}
    \caption{ The behavior of the occurrence rate of our classified CEMP candidate stars, as a function of metallicity ([M/H]) provided by \citet{Andrae2023}. Specifically, we show the fraction of stars with [M/H] < $x$ that are classified as CEMP out of the total number of BP/RP spectra that we classify in a given metallicity bin. We also plot the metallicity distribution of our CEMP candidate sample in grey.  We note this plot is only to explore the properties of our sample, and is not meant as a measure of the true occurrence rate of CEMP stars. The CEMP fraction increases with decreasing metallicity, consistent with previous results from high-resolution samples.}  
    \label{fig:binary}
\end{figure}

\begin{figure*}

    \includegraphics[width=\linewidth]{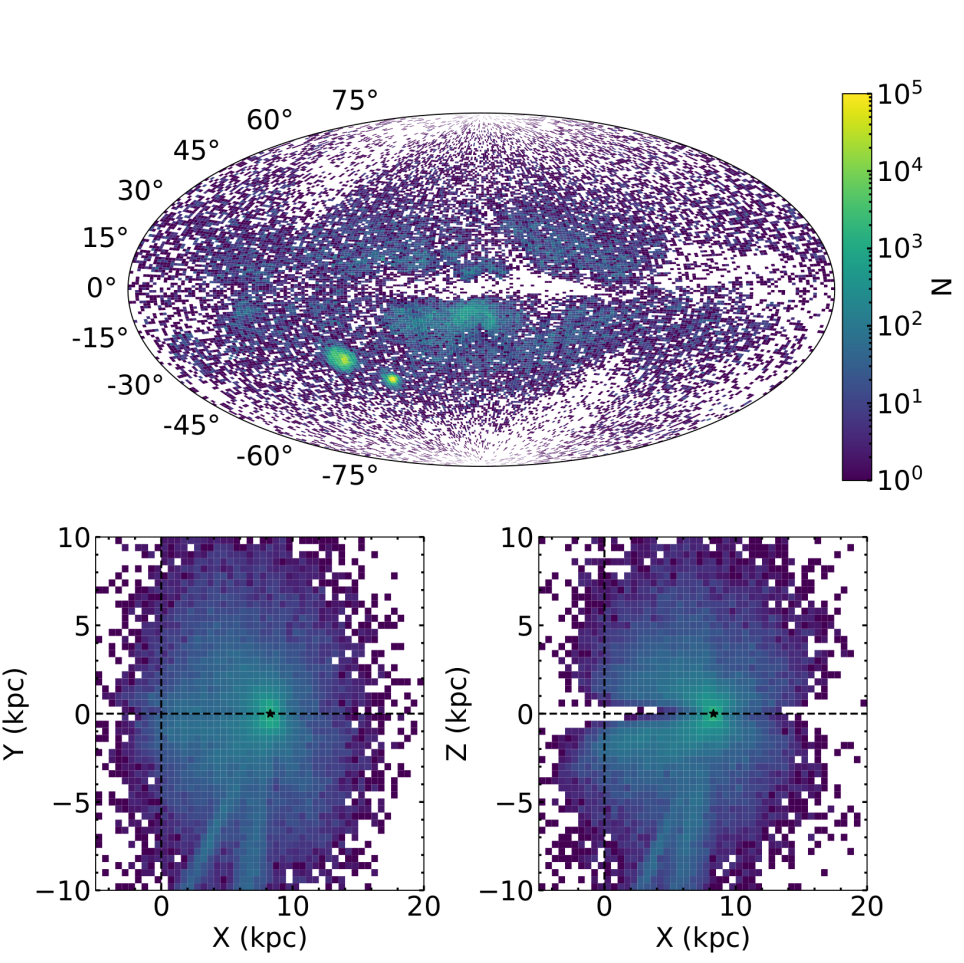}
    \caption{ The Galactic distribution of our CEMP candidate sample. Specifically, in the top panel we show the sky projected distribution of the sample in Galactic coordinates, $l$ and $b$, where ($l$,$b$) = (0,0)$^{\circ}$ is the line-of-sight towards the Galactic center. The bottom left panel shows the distribution of stars in the Galactic coordinates  X and Y. The bottom right panel shows the distribution in the Galactic X and Z coordinates. The Galactic center is located at (0,0,0) kpc, while the Sun is at (8.3,0.0) kpc. The LMC and SMC are clear features in the sky projected distribution. }
    \label{fig:map}
\end{figure*}

\begin{figure*}

    \includegraphics[width=\linewidth]{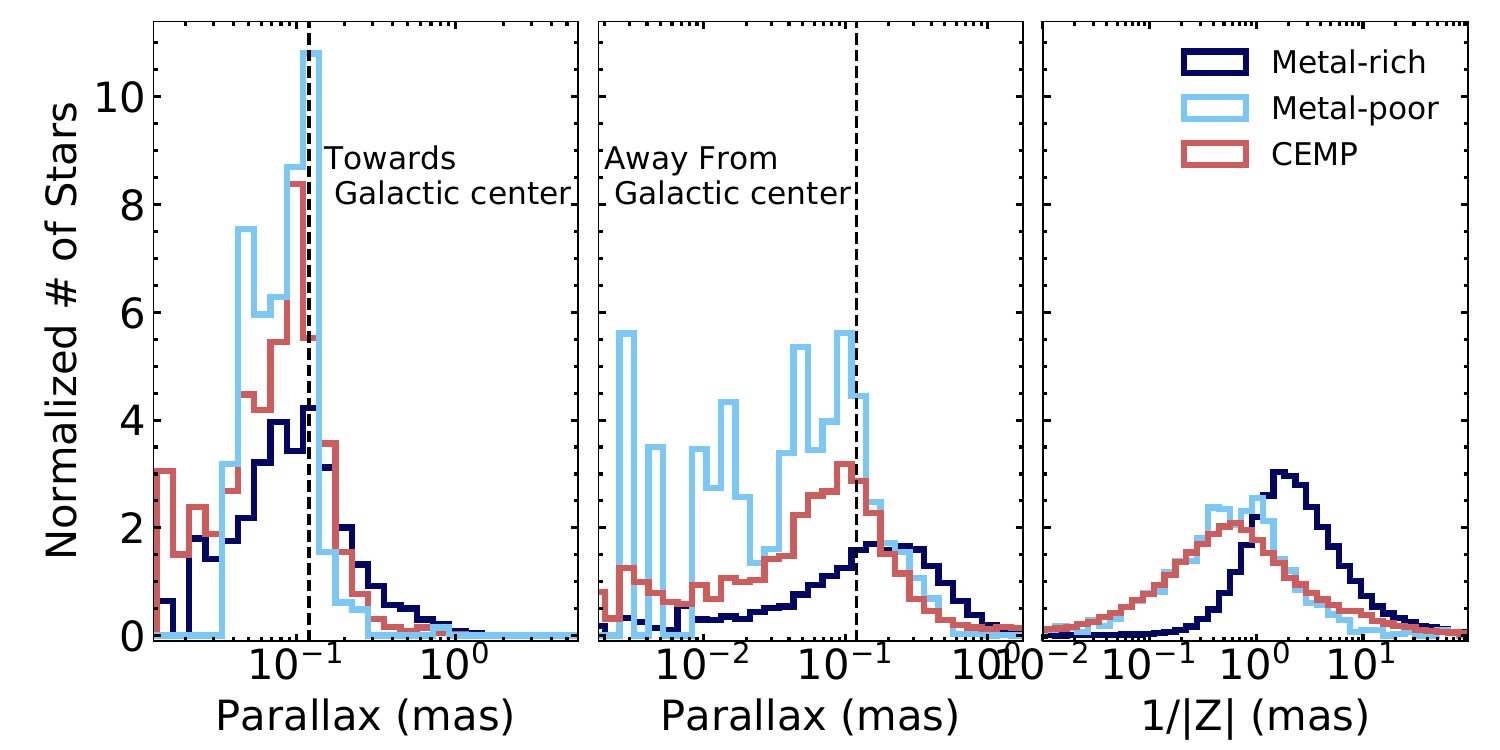}
    \caption{ The parallax distribution of our CEMP candidate sample  (red), compared to the metal-rich (dark blue) and metal-poor  (light blue) carbon-normal sample. Specifically, in the left panel, we show the parallax distributions for stars towards the Galactic center with |$l$| < 10$^{\circ}$ (or 350$^{\circ}$ < $l$ < 10$^{\circ}$) and |$b$| < 10$^{\circ}$. The middle panel shows the parallax distributions for stars towards the Galactic anti-center with 170$^{\circ}$ < $l$ < 190$^{\circ}$ and |$b$| < 10$^{\circ}$. We also mark the parallax of the Galactic center (0.12 mas) with a black vertical dashed line in both the left and middle panels. The right panel shows the inverse distance from the Galactic plane (|Z|) calculated by dividing the parallax by $\rm{sin}(b)$. The CEMP candidate stars generally follow the distribution of metal-poor stars.}
    \label{fig:dist}
\end{figure*}

Out of the $\approx$180 million stars that we classify, we find 58,872 CEMP candidate stars. This is the largest, homogeneously identified sample of CEMP candidate stars to date. In this section, we briefly investigate a few properties of this sample, including their metallicity distribution and Galactic distribution. 

Figure \ref{fig:cmd_f} shows the color-magnitude diagram of our CEMP candidate sample compared to a random sample of stars classified as carbon-normal stars of the same size. We show the CEMP candidate sample in a gradient from dark blue to beige, where dark blue shows the densest area of stars. The carbon-normal sample is shown as contour lines, where the highest density is shown with white contour lines; lower density areas have black contour lines. In general, the color-magnitude distribution of the CEMP candidate and carbon-normal samples are similar, in that the highest density occurs where we expect bright dwarf and turn-off stars to reside ($G\approx$ 4-6). However, one key difference occurs on the giant branch ($G<4$). In general, the CEMP candidate giant stars are bluer than the carbon-normal giant stars. There is a clear red clump feature in the carbon-normal giant star distribution (black contour) at $G \approx$ 1 and \bprp$\approx$ 1.6, but this does not appear in the CEMP distribution. This is consistent with expectations that the CEMP candidate sample is more metal poor than the carbon-normal sample, given that metal-poor stars will become blue-horizontal branch (BHB) after the RGB phase rather than red-clump stars like their more metal-rich counterparts. Another interesting feature of the CEMP distribution is a distinct group of dwarf stars at 1.25 $<$\bprp$<$ 1.50 and $G<5$. This is similar to the distribution of dwarf stars in the training sample (see the left panel of Figure \ref{fig:train}). This group appears to be brighter than the typical main sequence stars in that range of \bprp. It is possible that this may be due to an unresolved binary companion or a reddening effect from strong molecular absorption feature in the blue for the corresponding \teff\ range. However, higher resolution follow-up is required to confirm these hypotheses.  

Figure \ref{fig:binary}  shows the metallicity distribution of our sample of CEMP candidate stars in a grey histogram, using data-driven results from \citet{Andrae2023}. In dark blue, we also show the ratio of the number of stars classified as CEMP to the total number of BP/RP spectra we analyzed for [M/H] < $x$. In other words, the dark blue line gives the fraction of CEMP candidate stars for stars with [M/H] $<$ x. It is important to note that this plot is only meant to explore the trends in our sample, and not meant as a measurement of the true occurrence rate of CEMP stars. With the uncertain metallicities, undefined selection function, and low completeness percentage, our sample currently cannot provide a robust estimate of the occurrence rate as a function of metallicity. However, the overall trend is consistent with previous work \citep{Lucatello2006,Lee2013,Placco2014,Yoon2018,Arentsen2022} in that  the occurrence rate increases with decreasing metallicity. Unexpectedly, we find that 57\% of our CEMP candidate sample has [M/H] estimates from \citet{Andrae2023} higher than --1, indicating that they are not metal poor. However, given that the [M/H] estimates are data-driven and CEMP stars are outliers, it is unlikely that the [M/H] values are robust for CEMP stars. It is likely that they are, in fact, over-estimated given the carbon enhancement, and the true [Fe/H] would be much lower. For example, using the 17 stars that have $p$(CEMP)$>$ 0.5 in our catalog and also [C/Fe] $>$ +0.7 measured from APOGEE, we find an average [Fe/H] = --1.39, with minimum [Fe/H] = --2.23 and maximum [Fe/H] = --0.82. On the other hand, the average [M/H] given by \citet{Andrae2023} for the same stars is [M/H] = --0.03, with minimum [M/H] = --0.51 and maximum [M/H] = +0.49. Comparing with the SEGUE training/testing sample, we find that the [M/H] given by \citet{Andrae2023} is over-estimated compared to [Fe/H] by an average of +0.88 dex for stars with [C/Fe] $>$ +0.7. The over-estimation increases with increasing [C/Fe] in that stars with [C/Fe] $>$ +4 have [M/H] $-$ [Fe/H] $>$ +3 dex.

Figure \ref{fig:map} shows the Galactic distribution of the CEMP candidate sample. Here we use the geometric distances from \citet{Bailer-Jones2021}. We choose not to use distances that are calculated using photometry, since it is likely that the carbon enhancement will bias these results. The top panel simply shows the sky-projected distribution in Galactic coordinates, $l$ and $b$, which is independent of the parallax. Here, the Galactic center is along the line-of-sight towards ($l$,$b$) = (0,0)$^{\circ}$. We also show the distribution of our CEMP candidate sample in Galactic X and Y coordinates in the bottom right panel, while the bottom left panel shows the distribution in Galactic X and Z coordinates. The Galactic center is located at (0,0,0) kpc with the Sun located at (8.3,0,0) kpc \citep{Reid2014}.

The CEMP candidate sample is spread throughout the Galaxy in a halo-like distribution. However, towards the inner Galaxy at low |$b$| there is an under-density of stars, likely caused by extinction. The Large and Small Magellanic Clouds (LMC and SMC) stand out as clear features in the sky projected distribution, with peaks in the number density of CEMP candidate stars. This is consistent with previous work that has identified thousands of carbon-rich (post-)AGB stars in the Magellanic clouds \citep{Rebeirot1993,Kontizas2001}. Similarly, the over-density of CEMP candidate stars towards the Galactic center at negative $b$ is likely associated with the Sagittarius dwarf galaxy. However, further work on the dynamics of these stars is required to tag them to a specific Galactic component. Additionally, spectroscopic observations are required to calculate evolutionary corrections, and determine which of these stars are potentially natal CEMP-no stars, rather than (post-)AGB carbon stars or CEMP-$s$ stars. Impacts of the Gaia DR3 selection function can also be seen in the sky projected distribution at high |$b$| with sweeping under-density features \citep[see Figure 29 in ][]{DeAngeli2022}. 

Consistent with a kinematically hot Galactic population, our CEMP candidate sample is extended to large distances from the Galactic center (see Figure \ref{fig:map}). It is important to note that some of the largest distances may be unreliable, given that the fractional parallax uncertainty for faint stars can be quite large. Given that the majority of BP/RP spectra released in DR3 have $G \approx$ 17.6, and assuming an absolute $G$ magnitude of $-2.5$ (roughly the tip of the RGB), we expect to have detected CEMP stars at distances of up to $\approx$ 30 kpc.  We find that 8,707 stars in our CEMP sample have parallaxes corresponding to distances from the Galactic center $>$ 30 kpc. However, the Galactic prior used in \citet{Bailer-Jones2021} brings most of these to within $\approx$ 10 kpc.

 We also investigate the relative parallax distribution of CEMP candidate stars (red), compared to carbon-normal metal-rich stars (dark blue) and carbon-normal metal-poor stars (light blue) in Figure \ref{fig:dist}. Specifically, we use the metallicity estimates from \citet{Andrae2023} with [M/H] $>$ --1 defined as metal rich and [M/H] $<$ --1 as metal poor. The left panel shows the parallax distribution of CEMP candidate stars towards the Galactic center by choosing stars with |$l$| < 10$^{\circ}$ and |$b$| < 10$^{\circ}$. We find that the distribution of CEMP candidate stars peaks at a parallax consistent with that of the Galactic center (0.12 mas). The distribution of metal-rich and metal-poor carbon-normal stars also peaks at this parallax, but the metal-rich stars have a stronger tail towards large parallaxes, i.e., closer distances to the Sun. The middle panel shows the parallax distributions of CEMP candidate and carbon-normal stars towards the Galactic anti-center with |$l$| < 170$^{\circ}$ and |$b$| < 10$^{\circ}$. We find that the parallax distribution of CEMP candidate stars and carbon-normal metal-poor stars peaks at a parallax of $\approx$ 0.1 mas. Given that the average parallax precision for the faintest stars in our sample ($G \approx 17.5$) is $\approx$ 0.1 mas \citep{Lindegren2021}, this is consistent with CEMP stars peaking at distances $\geq$ 10 kpc. The carbon-normal metal-rich stars, on the other hand, peak at large parallaxes, indicating they are generally closer to the Sun than CEMP stars. This is consistent with previous work, which found the frequency of CEMP stars to increase with increasing distance from the Sun \citep{Frebel2006,Carollo2012,Lee2013,Lee2017,Yoon2018}.  In the right panel, we show the inverse of the distance from the Galactic plane (1/|Z|), which is calculated by dividing the parallax by $\rm{sin(}b)$.  Again, we find that the CEMP candidate and metal-poor carbon-normal stars follow the same trends, and are generally farther from the Galactic plane than the carbon-normal stars.   In general, we find that the CEMP candidate stars are more distant from the Sun than the carbon-normal stars, consistent with a Galactic halo population. Further work looking at the rate of carbon enhancement for metal-poor stars as a function of Galactic position is required to determine if CEMP stars have different origins than carbon-normal metal-poor stars. 

\section{Summary} 

The origins of CEMP stars are poorly understood, even though they comprise $\approx$ 30\% of stars with [Fe/H] < --2 \citep{Lucatello2006,Lee2013,Placco2014,Yoon2018,Arentsen2022}. A significant fraction of CEMP stars have enhancements in $s$-process elements, and are called CEMP-$s$ stars \citep{Beers2005}. These stars are thought to receive their over-abundant carbon and $s$-process elements from a mass-transfer event with their binary companion, which has evolved to or past the AGB \citep{Lugaro2012,Placco2013}. On the other hand, CEMP stars without neutron-capture enhancements,  the CEMP-no stars, are thought to have been primarily enriched by material from the first generations of stars \citep{Umeda2003,Chiappini2006,Meynet2006,Nomoto2013,Tominaga2014}. However, there are many remaining questions about these unique stars, including why they seem to be less frequent in the central regions of our Galaxy, where we expect the highest concentration of ancient stars to reside \citep{Howes2014,Howes2015,Howes2016,Arentsen2021,Lucey2022}. As suggested by \citet{Yoon2019}, this dearth of CEMP stars may be caused by the dilution of CEMP stars in more massive subsystems (e.g., dwarf galaxies) with prolonged star-formation histories, which could be the origin of metal-poor stars in the inner Galaxy. However, given that the discrepancy in the CEMP fraction of the inner Galaxy is highest for [Fe/H] $>$ -2.5, \citet{Arentsen2021} argue that it is due to a lower rate of CEMP-$s$ stars caused by a lower binary fraction in the inner Galaxy.

In this work, we leverage the data from the all-sky \gaia\ survey to identify 58,872 CEMP candidates. Specifically, we use the $\sim$180 million BP/RP spectra made available in \gaia\ DR3. Using the \texttt{XGBoost} algorithm for classification, we achieve a completeness of 26\% and a contamination rate of 12\%. When comparing to high-resolution catalogs of CEMP stars \citep{zepeda2023}, we find that we positively identify 60-68\% of the previously known CEMP stars in the \gaia\ DR3 BP/RP data.  We ensure that the \texttt{XGBoost} algorithm matches our physical intuition and primarily performs the classification using spectral features that are correlated with the carbon abundance. 

We briefly investigate a few of the properties of our CEMP candidate sample, including the metallicity distribution and the Galactic spatial distribution. As expected, the CEMP fraction increases with decreasing metallicity. In general, we find that the CEMP candidate stars tend to follow the distribution of metal-poor carbon-normal stars, and that they are farther from the Sun than metal-rich carbon-normal stars. 

In future work, we plan to look at the orbital properties of a subset of these stars, as well as the rate of $s$-process enhancement. We plan to perform medium- and high-resolution spectroscopic follow-ups of many of these targets in order to confirm our contamination rate, measure radial velocities where needed, derive their dynamical properties (see, e.g., \citealt{Dietz2020,Dietz2021}), and identify chemo-dynamical groups (see, e.g., \citealt{zepeda2023}), and determine their neutron-capture abundances. We plan to specifically follow-up targets towards the central region of the Galaxy where the number of known CEMP stars is much lower. Following future \gaia\ releases, which will include many millions more BP/RP spectra, we expect to continue this work and again increase the number of CEMP candidate stars.

\appendix
\section{Online Table of CEMP Candidates}
We show a section of the available online table in Table \ref{tab:table}. The online table includes the source IDs for all 182,815,672 \gaia\ DR3 objects with BP/RP spectra within our color cuts. We also provide the inferred probability of being a CEMP star for each star.

\begin{table}
\caption{CEMP Probability Catalog.}
\label{tab:table}
\begin{tabular}{cc}
\hline\hline
Source ID & $p$(CEMP) \\
\hline
909506570173184 & 0.88\\
1012963742461952 & 0.00\\
1200018158226816 & 0.91 \\
1235889725071360 & 0.99 \\
1452184278048512 & 0.00 \\
7683563349186048 & 0.00\\
7746888346949120 & 0.00 \\
8731294851221504 & 0.63 \\
8779982600469632 & 0.99 \\
8797952743613440 & 0.52 \\
... & ...\\
\hline
\end{tabular}
\flushleft{A sample of the provided online catalog of CEMP probabilities. We provide the \gaia\ DR3 source ID for each star and the corresponding probability of being a CEMP star ($p$(CEMP)). }
\end{table}

\section*{Acknowledgements}
{\small 
This project was developed in part at the Gaia F\^ete, held at the Flatiron institute Centre for Computational Astrophysics in June 2022. We would like to thank David Hogg and Jason Hunt for their helpful discussions and computational resources. We also thank an anonymous referee for a careful review of this paper, which has improved the clarity of its presentation.

This material is based upon work supported by the National Science Foundation Graduate Research Fellowship under Grant No. 000392968. KH acknowledges support from the National Science Foundation grant AST-1907417 and AST-2108736 and from the Wootton Center for Astrophysical Plasma Properties funded under the United States Department of Energy collaborative agreement DE-NA0003843. This work was performed in part at the Simons Foundation Flatiron Institute's Center for Computational Astrophysics during KH's tenure as an IDEA Fellow. Y.S.T. acknowledges financial support from the Australian Research Council through DECRA Fellowship DE220101520. Work at Argonne National Laboratory was supported by the U.S. Department of Energy, Office of High Energy Physics. Argonne, a U.S. Department of Energy Office of Science Laboratory, is operated by UChicago Argonne LLC under contract no. DE-AC02-06CH11357. N.R acknowledges the Laboratory Directed Research and Development (LDRD) funding from Argonne National Laboratory, provided by the Director, Office of Science, of the U.S. Department of Energy under Contract No. DE-AC02-06CH11357.  T.C.B. and Y.S.L. acknowledge partial support for this work from grant PHY 14-30152; Physics Frontier Center/JINA Center for the Evolution of the Elements (JINA-CEE), and OISE-1927130: The International
Research Network for Nuclear Astrophysics (IReNA), awarded
by the US National Science Foundation.  Y.S.L. also acknowledges support from the National Research Foundation (NRF) of Korea grant funded by the Ministry of Science and ICT (NRF-2021R1A2C1008679).

This work has made use of data from the European Space Agency (ESA) mission{\it Gaia}  (\url{https://www.cosmos.esa.int/gaia}), processed by the {\it Gaia} Data Processing and Analysis Consortium (DPAC, \url{https://www.cosmos.esa.int/web/gaia/dpac/consortium}). Funding for the DPAC has been provided by national institutions, in particular the institutionsparticipating in the {\it Gaia} Multilateral Agreement.

Funding for SDSS-III has been provided by the Alfred P. Sloan Foundation, the Participating Institutions, the National Science Foundation, and the U.S. Department of Energy Office of Science. The SDSS-III website is http://www.sdss3.org/.

SDSS-III is managed by the Astrophysical Research Consortium for the Participating Institutions of the SDSS-III Collaboration including the University of Arizona, the Brazilian Participation Group, Brookhaven National Laboratory, Carnegie Mellon University, University of Florida, the French Participation Group, the German Participation Group, Harvard University, the Instituto de Astrofisica de Canarias, the Michigan State/Notre Dame/JINA Participation Group, Johns Hopkins University, Lawrence Berkeley National Laboratory, Max Planck Institute for Astrophysics, Max Planck Institute for Extraterrestrial Physics, New Mexico State University, New York University, Ohio State University, Pennsylvania State University, University of Portsmouth, Princeton University, the Spanish Participation Group, University of Tokyo, University of Utah, Vanderbilt University, University of Virginia, University of Washington, and Yale University.

This research made use of Astropy,\footnote{http://www.astropy.org} a community-developed core Python package for Astronomy \citep{astropy:2013, astropy:2018}. Other software used includes \textit{IPython} \citep{ipython}, \textit{matplotlib} \citep{matplotlib}, \textit{numpy} \citep{numpy}, \textit{scikit-learn} \citep{scikit-learn} and \textit{scipy} \citep{scipy}.
}

\section*{Data Availability}
The data underlying this article are publicly available through the \textit{Gaia} Archive.\footnote{\url{https://gea.esac.esa.int/archive/}} The generated catalog of CEMP candidate stars will be available in the online supplementary material and CDS upon acceptance. The \texttt{XGBoost} model and all code will be made available on Github upon acceptance.

\bibliography{bibliography}
\bsp	% typesetting comment
\label{lastpage}
\end{document}